\RequirePackage[2024-06-01]{latexrelease}
\documentclass[%
 reprint,
superscriptaddress,
 amsmath,amssymb,
 aps
]{revtex4-2}
\ExplSyntaxOn
\DeclareRobustCommand{\MakeTextUppercase}[1]{\text_uppercase:n{#1}}
\DeclareRobustCommand{\MakeTextLowercase}[1]{\text_lowercase:n{#1}}
\ExplSyntaxOff

\usepackage{graphicx}
\usepackage{dcolumn}
\usepackage{bm}
\usepackage{xcolor}
\usepackage{multirow}
\usepackage{hhline}
\usepackage{makecell}

\usepackage{tabularx}
\usepackage{array}
\newcolumntype{C}[1]{>{\centering\arraybackslash}p{#1}}

\usepackage{color}
\usepackage{float}
\usepackage{placeins}

\begin{document}

\preprint{APS/123-QED}

\title{\Large \textbf{Combining Quasiparticle Self-Consistent $GW$ and Machine-Learned DFT+$U$ to Assess Half-Metallicity in Co- and Ni-Based Heuslers}}

\author{Zefeng Cai}
\affiliation{Department of Materials Science and Engineering, Carnegie Mellon University, Pittsburgh, PA 15213, USA}
\author{Malcolm J. A. Jardine}
\affiliation{Department of Materials Science and Engineering, Carnegie Mellon University, Pittsburgh, PA 15213, USA}
\author{Maituo Yu}
\affiliation{Department of Materials Science and Engineering, Carnegie Mellon University, Pittsburgh, PA 15213, USA}
\author{Chenbo Min}
\affiliation{Department of Materials Science and Engineering, Carnegie Mellon University, Pittsburgh, PA 15213, USA}
\author{Jiatian Wu}
\affiliation{Department of Materials Science and Engineering, Carnegie Mellon University, Pittsburgh, PA 15213, USA}
\author{Hantian Liu}
\affiliation{Department of Materials Science and Engineering, Carnegie Mellon University, Pittsburgh, PA 15213, USA}
\author{Derek Dardzinski}
\affiliation{Department of Materials Science and Engineering, Carnegie Mellon University, Pittsburgh, PA 15213, USA}
\author{Christopher J. Palmstr{\o}m}
\affiliation{Materials Department, University of California-Santa Barbara, Santa Barbara, CA 93106, USA}
\affiliation{Department of Electrical and Computer Engineering, University of California-Santa Barbara, Santa Barbara, CA 93106, USA}
\author{Noa Marom}
 \email{Electronic mail: nmarom@andrew.cmu.edu}
\affiliation{Department of Materials Science and Engineering, Carnegie Mellon University, Pittsburgh, PA 15213, USA}
\affiliation{Department of Chemistry, Carnegie Mellon University, Pittsburgh, PA 15213, USA}
\affiliation{Department of Physics, Carnegie Mellon University, Pittsburgh, PA 15213, USA}

\date{\today}

\begin{abstract}
Half-metallic Heusler compounds are of significant interest for spintronics. For device fabrication, compounds that can be epitaxially grown on III-V semiconductors are particularly attractive. We present a first-principles investigation of four Co-based and two Ni-based Heusler compounds that are lattice-matched to InAs. The results of density functional theory (DFT) using semi-local and hybrid functionals are compared to quasiparticle self-consistent $GW$ (QP$GW$). We also consider DFT with machine-learned Hubbard $U$ corrections [npj Computational Materials 6, 180 (2020)] with a new Bayesian optimization (BO) objective function to determine the $U$ values that yield the closest agreement with the QP$GW$ band structure and magnetic moments. We find that DFT+$U$(BO) can adequately reproduce the key QP$GW$ features in most cases. Our results reveal a strong method dependence of the degree of spin polarization at the Fermi level and, in some cases, even the dominant spin channel (majority or minority). Of the materials studied here, Co$_2$TiSn and Co$_2$ZrAl are the most likely to be half-metals.

\end{abstract}

\maketitle

\section{\label{sec:level1} Introduction}

The Heuslers are a large family of ternary compounds  with the formula X$_2$YZ, where X and Y are transition metals or rare earth elements and Z is a $p$-block element \cite{1912_Knowlton_Heuslers}. The sustained interest in Heuslers has been driven by their useful properties for a wide array of applications \cite{felser2015heusler, graf2011simple, felser2015basics, farshchi2013spin, wollmann2017heusler, graf2010heusler, hirohata2022heusler}.  Magnetic Heusler compounds can serve as a ferromagnetic contact in magnetic tunnel junctions and magnetic spin valves \cite{peterson2016spin, sahoo2016compensated}. This can significantly improve the tunneling magnetoresistance (TMR) \cite{kammerer2004co, okamura2005large}, which is a key metric for efficient magnetic transport in a hard drive or a magnetic random access memory (MRAM) chip \cite{elphick2021heusler, palmstrom2016heusler}. Furthermore, the high spin polarization displayed by some Heuslers would make them ideal spin contacts to the semiconductor in spin-field effect transistors (FETs) \cite{datta1990electronic}, spin light emitting diodes \cite{holub2007electrical}, and spin-lasers \cite{bhattacharya2010quantum}. More recently, some members of the Heusler family have been found to be topological insulators \cite{yan2017topological, chadov2010tunable}. Here, we focus on Heusler compounds that may exhibit half-metallicity \cite{1983_Groot_OGHeuslerTheory_PRL, 1995_Ishida_HeuslerLDAHalfMetal_JPhysSocJapan, galanakis2008ferrimagnetism, felser2015heusler, 2025_Nishihaya_PRB}, meaning that one spin channel is metallic and the other is semiconducting.  Half-metals are of interest for spintronics thanks to the promise of achieving close to 100\% spin polarization at the Fermi level.

Chemically ordered films and atomically sharp, defect-free interfaces are desirable for spintronic devices because the preservation of half-metallicity in Heusler alloys and spin-dependent transport across interfaces are highly sensitive to disorder, interfacial roughness, and impurities \cite{elphick2021heusler, palmstrom2016heusler, orgassa1999first, van2000epitaxial, sakuraba2005fabrication, adelmann2005effects, marukame2007highly, schultz2009spin, guillemard2020issues, de2002hybrid, parkin2004giant, yuasa2004giant, velev2008interface, wang2010structural, hirohata2022interfacial}. 
In particular, deviations from ideal L2$_1$ ordering in full-Heusler compounds have been shown to strongly suppress spin polarization \cite{elphick2021heusler} and increase magnetic damping \cite{guillemard2020issues}. Interfacial defects and disorder may introduce spin-flip scattering that reduces tunneling magnetoresistance and spin injection efficiency \cite{de2002hybrid, wang2010structural, hirohata2022interfacial}. It has also been reported that half-metallicity may be degraded near surfaces and interfaces \cite{2003_Picozzi_SpinInterfaceHeuslerSM_JApplPhys,khosravizadeh2009first,galanakis2007spin,2005_Galanakis_NiMnXInPGaAsInterface_PRB,2002_Galanakis_HalfFullHeuslerSurfaces_JPhysACondensMatter,2004_Jenkins_HalfMetalSurface_PRB,2001_Jenkins_InterfaceStatesHalfMetalNiMnSb001_SurfSicLet}. It has been found that the properties of the interface are governed by the local magnetic ordering and electronic structure of both the Heusler compound and the semiconductor \cite{palmstrom2016heusler, rath2018reduced, heischmidt2022first}.

These considerations motivate the search for Heuslers that can form epitaxial interfaces with appropriate substrates via growth techniques that enable precise stoichiometric and structural control.
Ordered Heusler films can be fabricated by high temperature annealing when in contact with thermodynamically stable metals or insulating oxides \cite{sakuraba2005huge, tezuka2006tunnel}. However, growing Heuslers on reactive semiconductors requires the use of molecular-beam epitaxy (MBE) \cite{hirohata2006heusler} with individual elemental sources. Heusler compounds have been grown on various III-V semiconductors with a compatible FCC lattice. Thanks to the large compositional space of the Heusler family, which may be further expanded by substitutional alloying, it is possible to find lattice matched Heuslers for substrates with a wide range of lattice parameters including GaAs, InP, InAs, GaSb, and InSb \cite{palmstrom2016heusler,2011_Alijani_HeuslerCoFeMnXExpTheory_PRB,2013_Gao_CoFeCrXGGAU_JAlloyComp,2005_Valerio_DepositionExpCo2MnXSn_AppliSurfSci}. The growth of epitaxial Heusler/III-V interfaces is nontrivial because the elements comprising the Heusler (such as Mn) may diffuse into and react with the III-V semiconductor at the interface \cite{law2024formation, 2025_Nishihaya_PRB, singh2006interface}, or conversely, elements from the semiconductor may out-diffuse into the Heusler film \cite{hamaya2022semiconductor}, requiring the use of barrier layers. Here, we focus on Heusler compounds predicted to have good lattice matching with InAs. InAs is attractive for spintronics and quantum computing applications due to its high electron mobility, large effective $g$-factor, strong spin-orbit coupling, and the absence of a Schottky barrier at the interface with various metals \cite{zhang2023mobility, prabhakar2011manipulation, takase2017highly, feng2016schottky}. Because GaSb has a close lattice parameter to InAs, the materials considered here are also matched to GaSb. Considering the high cost of MBE, computer simulations can help select the most promising candidate materials for growth experiments. 

There is a long history of first principles studies of Heusers and related half-Heuslers starting from the seminal paper by de Groot \textit{et al.}, which  established the concept of half-metallicity in half-Heusler compounds  \cite{1983_Groot_OGHeuslerTheory_PRL}. Subsequent experimental work confirmed half-metallic behavior in materials such as NiMnSb \cite{1986_Hanssen_positronNiMnSbExp_PRB,1986_Hanssen_positronNiMnSbTheory_PRB,kirillova1995interband}. This discovery motivated extensive computational investigations, predominantly using density functional theory (DFT) with local or semi-local exchange–correlation functionals \cite{1995_Ishida_HeuslerLDAHalfMetal_JPhysSocJapan,2002_Galanakis_HeuslerGapLDAKKR_PRB,2006_Galanakis_ElectronicMagneticHeusler_JPhysApplPhys}. In recent years, high-throughput computational screening has been employed to discover and design new Heusler candidates for diverse applications \cite{2012_Roy_PRL, yan2014half, sanvito2017accelerated,rotjanapittayakul2018search,ma2017computational,ma2018computational,jiang2021review,jiang2022high,jiang2023high, 2025_Hilgers_PRM}. One limitation of such studies is that the predicted properties of Heusler compounds (and related classes of materials), including band gaps and magnetic properties, are highly sensitive to the choice of approximation for the exchange–correlation functional. Local and semi-local functionals may fail to reliably describe $d$-electron magnetism and half-metallicity  in Heuslers and related materials \cite{2019_Nawa_DFTUCoBasedHeusler_RSCAdv,2006_Kendpal_LDAUNeedCorrelation_PRB,faleev2017unified,faleev2017origin,2017_Faleev_HeuslerMagneticTunnelLDA_PRM,1997_Anisimov_LDAUStronglyCorrelatedMatterials_JPhysACondensMatt,2008_Kobayashi_GWExampleNiMnCompound_PRB,heischmidt2022first}. The deficiencies of (semi-)local DFT functionals may be attributed to a combination of the self-interaction error (SIE) and an inadequate treatment of strong correlation effects, both particularly pronounced for localized $d$- and $f$-electrons \cite{SIE, mori2008localization, cohen2008insights, kotliar2006electronic}.

Hybrid functionals, in which a fraction of exact (Fock) exchange is mixed with semi-local exchange and correlation, often produce more accurate band gaps and band structures  \cite{2017_Shi_HSEBanfGapsHalfHeusler_JApplPhys,2011_Bilc_Fe2VAlHybrid_PRB}. However, hybrid functionals may overestimate exchange splitting and magnetic moments in transition-metals \cite{2018_Jana_PropertiesMagMomFunctional_JChemPhys,paier2006screened,2019_Fu_DFTMEthodsMagnetismTransitionMetals_PRB,2014_Janthon_TransitionMetalsFunctionals_JCTC,2018_Zahedifar_HalfHeuslerHSEGW_PRB}, which are the most common components in Heusler compounds, and may inaccurately describe spectral features near the Fermi level \cite{2016_Gao_ApplicabilityHybrid_SolidStatComm,coulter2013limitations}. Further improvement may be achieved by going beyond DFT.
Many-body perturbation theory in the $GW$ approximation \cite{golze2019gw}, where $G$ represents the Green’s function and $W$ the screened Coulomb interaction, provides a more reliable description of correlated and magnetic materials \cite{2021_Acharya_GWOthersExampleCompare_npjcompumats,2015_Jiang_UsingGWStronglyCorrelatedStartingPoint_IntJQuanChem,coulter2013limitations,yamasaki2003electronic,2008_Kotani_SpinWaveQPGWMagneticMat_JPhysCondensMatter}, including Heusler and half-Heusler compounds \cite{2018_Zahedifar_HalfHeuslerHSEGW_PRB,2023_Gurbuz_FunctionalCalcExchangeCorrel_JournApplPhys,2017_Tas_DFTGWSpinFilterHuesler_JourMagnetMagMatterials, 2025_Nishihaya_PRB}. In particular, the quasiparticle self-consistent $GW$ (QP$GW$) method \cite{2004_Faleev_QPGWInitial_PRL,Schilfgaarde_2006_QPGWOG_PRL,shishkin2007accurate} eliminates the dependence on the mean-field starting point, which is a known limitation of one-shot $G_0W_0$  \cite{2023_Cunningham_QSGWHatVariousMatt_PRB,kotani2009re,kotani2007quasiparticle,2022_Friedrich_SimpleMetalGWImportant_NanoMatt,ismail2017justifying}. However, the high computational cost of hybrid functionals, and even more so of $GW$, limits the applicability of these methods to relatively small systems and precludes their utilization for studies of large interface models. 

The DFT+$U$ method provides a balance between accuracy and efficiency by adding Hubbard $U$ corrections that introduce an orbital-dependent on-site Coulomb interaction to (semi-)local functionals  \cite{dudarev1998electron}. DFT+$U$ with an appropriate choice of the system-dependent $U$ parameter(s) has been widely applied to correlated materials with localized $d$ or $f$ orbitals, such as transition-metal compounds \cite{himmetoglu2014hubbard,2004_Zhou_TransitionMetalLDAURedox_PRB, aryasetiawan2006calculations, Mosey2008a, 2006_Kulik_PRL, jiang2010first, Harald_DFT+U}.  For Heusler compounds and related materials, introducing Hubbard $U$ corrections has been shown to improve the resulting electronic structure, including magnetic moments, spin polarization and band widths \cite{2006_Kendpal_LDAUNeedCorrelation_PRB,2007_Kandpal_LDAUHeuslerCalcProperties_JPhysDA_ApplPhys,2005_Wurmehl_HeuslerLDA_PRB,khosravizadeh2009first,2010_Nourmohammadi_Co2XSiPBE0Hybrid_SolidStateComm, 2013_Sasioglu_DFTUHalfMetalHeusler_PRB,2007_Fecher_LDAUBandGap_JPhysDApplPhys,mellah2022exchange, 2015_Baral_ExpCo2MnSbPESDFT_JournAlloyComp, sasioglu2025itinerant}. We have developed a method of machine learning the $U$ values for a given material using Bayesian optimization (BO) with an objective function that aims to reproduce as closely as possible the band structure obtained from a higher-level reference method such as a hybrid functional or QP$GW$ \cite{yu2020machine}. The DFT+$U$(BO) method has enabled us to perform large-scale simulations of surfaces and interfaces \cite{yang2021first,yu2021dependence,yang2022electronic, 2023_Jardine_TrilayerTunnelBarrier_ACS, jardine2025first}, including Heusler/ III-V interfaces \cite{heischmidt2022first}, which would not have been possible otherwise.

Here, we investigate the electronic and magnetic properties of four Co-based and two Ni-based Heusler compounds, whose lattice parameters are matched to InAs (or GaSb), as summarized in Table \ref{tab:table1}. We compare the results obtained from QP$GW$ to DFT with the generalized gradient approximation (GGA) of Perdew, Burke, and Ernzerhof (PBE) \cite{1996_Perdew_GeneralizedGradientApproximation_PRL, PBE_erratum} and the range-separated hybrid functional of Heyd-Scuseria-Ernzerhof (HSE) \cite{Heyd2003,Heyd2003_erratum}, as well as PBE+$U$(BO). For DFT+$U$(BO), we have formulated a new objective function that includes the atomic magnetic moments in addition to the band structure, and we use QP$GW$ as the reference method. The optimal $U$ values obtained from the BO procedure enable PBE+$U$(BO) to reproduce the key qualitative features of the QP$GW$ electronic structure. We find that the choice of electronic structure method strongly influences the results. In some cases, different methods even provide \textit{qualitatively} different predictions for whether a certain material is a half-metal and whether the spin polarization at the Fermi level is up or down. Of the materials studied here, only Co$_2$TiSn is unanimously predicted by all methods to be a half metal. Co$_2$ZrAl is predicted by all methods except HSE to be a half metal. Co$_2$MnIn is predicted by QP$GW$ to have a high degree of minority spin polarization at the Fermi level. Our findings have significant implications for the reliability of high-throughput studies of magnetic Heuslers based on semi-local DFT.

\begin{table*}
\caption{Summary of the computed equilibrium lattice constant, total magnetic moment, spin polarization at the Fermi level, and effective $U$ values applied to the $d$ orbitals of the X and Y elements of each Heusler compound (a $U$ correction was not applied to the Z element). The results of different electronic structure methods are compared for the total magnetic moment and the spin polarization at the Fermi level. A negative spin polarization indicates that the minority spin channel is dominant at the Fermi level.}
\label{tab:table1}
\vspace{2pt}
\centering
\renewcommand{\arraystretch}{1.45}
\setlength{\extrarowheight}{1pt}
\setlength{\tabcolsep}{7.8pt}

\begin{tabular}{c|c|ccC{25pt}C{30pt}|ccC{26pt}C{33pt}|cc}
\hhline{=|=|====|====|==}
\multirow{2}{*}{\centering Material} &
\multirow{2}{*}{\centering $a_0$ (\AA)} &
\multicolumn{4}{c|}{Total Magnetic Moment ($\mu_B$)} &
\multicolumn{4}{c|}{Spin Polarization (\%)} &
\multicolumn{2}{c}{\centering $U$ (eV)} \\

& & PBE & HSE & QP$GW$ & PBE+$U$ & PBE & HSE & QP$GW$ & PBE+$U$ & X & Y \\
\hhline{-|-|----|----|--}

Co$_2$MnIn & 5.980 & 4.455 & 7.374 & 6.448 & 5.853 & 85.98  & -64.36 & -94.62 & -94.29 & 2.252  & 0.751 \\
Co$_2$MnSn & 5.984 & 5.030 & 6.295 & 5.853 & 5.257 & 98.68 & -82.28 & -66.67 & 94.71  & 1.512  & 1.652 \\
Co$_2$TiSn & 6.082 & 1.997 & 2.009 & 1.996 & 1.997 & 100.00 & 100.00 & 100.00 & 99.98  & 3.534  & 3.654 \\
Co$_2$ZrAl & 6.063 & 0.999 & 1.304 & 1.000 & 0.999 & 100.00 & -32.83 & 100.00 & 100.00 & 1.331  & 9.379 \\
Ni$_2$MnSb & 6.059 & 4.044 & 4.462 & 4.662 & 4.620 & -30.77 & 14.94  & -43.03 & 7.85   & -1.512 & 5.155 \\
Ni$_2$MnSn & 6.056 & 4.108 & 4.613 & 4.780 & 4.466 & -67.52 & -65.63 & -69.93 & -56.95 & 2.392  & 2.352 \\

\hhline{=|=|====|====|==}
\end{tabular}
\end{table*}

\section{\label{sec:level2} Methods}

\subsection{DFT+$U$(BO)}
Previously, we had proposed Bayesian optimization as a practical approach for finding the optimal Hubbard $U$ parameters for a given material within the Dudarev formulation of DFT+$U$ \cite{dudarev1998electron}. The $U$ corrections can be applied to any (semi-)local functional. Here, we use PBE. The BO objective function is formulated to reproduce as closely as possible the results of a reference method ~\cite{yu2020machine}. Originally, the objective function included the first two terms of Eq.~\ref{Eq:objective}, \textit{i.e.,} the difference in the band gap, \(\Delta\)Gap (only used for semiconductors), and the difference in the band structure, \(\Delta\)Band, weighted by \(\alpha_1\) and \(\alpha_2\), respectively~\cite{yu2020machine}. For magnetic materials, we now introduce a third term, \(\Delta\)Mag, representing the difference in atomic magnetic moments, with an additional weight \(\alpha_3\):
\begin{equation}\label{Eq:objective}
\begin{aligned}
    {f(\vec{U})} = &-{\alpha_1}(\Delta \text{Gap})^2 \\
    &-{\alpha_2}(\Delta \text{Band})^2  \\
        &-{\alpha_3}(\Delta \text{Mag})^2
\end{aligned}
\end{equation}
where $\vec{U}$ = [$U^{1}$, $U^{2}$, \ldots, $U^{n}$] represents the vector of $U$ values applied to different elements and $U^{i} \in [-10, 10]$ eV. \(\Delta \text{Gap}\) is defined as:
\begin{equation}
\Delta \text{Gap}=E_{\text{g}}^{\text{ref}}- E_{\text{g}}^{\text{PBE+U}}
\end{equation}
 $\Delta \text{Band}$ is defined as
 an extension of the previous definition of the root mean squared difference between the PBE+$U$ and the reference band structure ~\cite{yu2020machine, 2017_Huhn_PRM}:
 \begin{equation}
\Delta \text{Band} = \sqrt{
\begin{aligned}
\frac{1}{N_E}\sum^{N_k}_{i=1}\sum^{N_b}_{j=1}w_{i}\Bigl(&(\epsilon_{\text{ref}}^{j}[k_i] - \epsilon_{\text{ref}, 0}^{j}) \\
- &(\epsilon_{\text{PBE+U}}^{j}[k_i] - \epsilon_{\text{PBE+U}, 0}^{j})\Bigr)^2
\end{aligned}
}
\end{equation}
Here, \(N_E\) is the total number of eigenvalues included in the comparison, \(N_k\) is the number of \(k\)-points, and \(N_b\) is the number of bands selected for comparison. The weight factor \(w_i\) reflects the relative importance of each \(k\)-point based on its contribution to the Brillouin zone (BZ) volume. For $k$-paths with an equal number of $k$-points, it is  normalized to ensure uniform sampling across the BZ:
\begin{equation}
\forall \, i \in I, \quad w_i = \frac{|\mathbf{k}_I|}{\sum_J |\mathbf{k}_J|} ,
\end{equation}
\(I\) and \(J\) are the index sets of \(k\)-points along the \(k\)-paths \(\mathbf{k}_I\) and \(\mathbf{k}_J\), respectively. \(|\mathbf{k}_I|\) denotes the length of the \(k\)-path \(\mathbf{k}_I\) in reciprocal space.
Here, $\epsilon_{\text{ref},0}^{j}$ and $\epsilon_{\text{PBE+U},0}^{j}$ denote reference energy levels defined for band $j$ in the reference method and in the PBE+$U$ calculation, respectively. These reference energies are introduced to account for the fact that absolute energy levels are not directly comparable between calculations performed with different exchange-correlation functionals or pseudopotentials; only relative energy differences are physically meaningful. The reference level can be chosen by the user. common choices are the respective vacuum level or the valence band maximum (VBM) in both methods. For metallic systems, the reference energy is typically taken to be the Fermi level of each calculation.
In the same vein, \(\Delta\)Mag is defined as the root mean squared error of the PBE+$U$ magnetic moments compared to the reference method:
\begin{equation}
\Delta \text{Mag} = \sqrt{\frac{1}{N_i}\sum^{N_i}_{i=1}(m_{\text{ref}}^{i} - m_{\text{PBE+U}}^{i})^2}
\end{equation}
where \(N_i\) is the total number of atoms, and \(m\) is the magnetic moment produced by each method. Including the magnetic term leads to faster and more robust convergence of the BO algorithm, as shown in the SI.

Previously, we used HSE as the reference method \cite{yu2020machine}. However, as discussed above, for the materials studied here, we consider QP$GW$ as a more reliable choice. 
For metallic Heusler compounds, the objective function coefficients are set to $\alpha_1 = 0.0$, $\alpha_2 = 0.5$, and $\alpha_3 = 0.5$, assigning equal weights to the band structure and magnetic moments.  $N_b$ is set to  16 with 8 bands above and 8 bands below the Fermi level. The parameter $\kappa$, which controls the balance between exploration and exploitation in the BO upper confidence bound acquisition function \cite{yu2020machine}, is set to 5.
For the Heuslers studied here, Hubbard $U$ corrections were applied to the $d$-orbitals of the transition-metal elements, X and Y, but not to the main-group element, Z. The BO algorithm was deemed converged if either (i) the difference in the objective function between two consecutive iterations was less than 0.0001 or (ii) the change in the interpolated optimum of the objective function over ten successive iterations was less than 0.005. 

The resulting $U$ values for each material are summarized in Table \ref{tab:table1} and BO convergence plots are provided in the SI. We note that the optimized quantity is the effective Hubbard parameter of the Dudarev formulation, $U_{\mathrm{eff}} = U - J$, rather than the bare on-site Coulomb repulsion $U$. Negative values of $U_{\mathrm{eff}}$ are therefore physically admissible: they correspond to the regime in which the on-site exchange $J$ exceeds the screened Coulomb repulsion $U$ \cite{1990_Micnas_NegativeUSuperconductivity_RMP, 2007_Hase_ValenceSkippingBaBiO3_PRB, 2009_Nakamura_LSDANegativeUIronSC_PhysicaC, 2012_Cococcioni_LDAUCorrelatedGroundStates_Juelich}. Negative $U_{\mathrm{eff}}$ values are appropriate for delocalized states, for which the exchange-correlation hole is overestimated by (semi-)local functionals \cite{2006_Persson_LDAUSICSemiconductors_BrazJPhys, yu2020machine}. We further note the origin of the large optimized $U_{\mathrm{Zr}}$ of Co$_2$ZrAl. The Zr $4d$ shell is nominally empty and lies well above the Fermi level, so the magnetic moments and the occupied bands are nearly insensitive to $U_{\mathrm{Zr}}$ [see Figure~S13(a) in the SI]. The optimization along $U_{\mathrm{Zr}}$ is therefore driven by the band-structure term alone, and the large value emerges naturally from improving the agreement of the unoccupied Zr-$4d$-derived conduction bands with QP$GW$ [Figure~S13(b)]. Including $U_{\mathrm{Zr}}$ is important because both the magnitude of the half-metallic gap and the transport in spintronic devices depend on the position of the unoccupied bands. A detailed comparison of optimizations with and without $U_{\mathrm{Zr}}$ is provided in the SI.

\subsection{Computational Details}

All calculations were performed using the Vienna Ab initio Simulation Package (VASP) with the projector augmented wave (PAW) method \cite{1993_Kresse_AbInitioOrigional_PRB,1999_Kresse_UltrasoftPseudopotentialsAumented_PRB,1994_Blochl_ProjectorAugmentedWaveMethod_PRB}. 
A plane-wave basis set was used with a kinetic energy cutoff of 400 eV and an energy convergence criterion of $10^{-6}$ eV.
Spin-orbit coupling (SOC) was used for all calculations \cite{2004_Mavropoulos_SOCBandGapHeusler_JourPhysCondensMatter}, including QP$GW$ (an analysis of the effect of SOC on the electronic structure is provided in the SI). 
The recommended  pseudopotentials provided by VASP (standard and $GW$) were used for all calculations. 
We included non-spherical contributions from the density gradient in the PAW spheres 
and applied a cutoff to the gradient field to restore the full lattice symmetry
in order to account for magnetic anisotropy. The Brillouin zone was sampled using the $\Gamma$-centered scheme. For QP$GW$ calculations, a $7 \times 7 \times 7$ $k$-point mesh was used. For PBE and HSE calculations a k-point mesh of $9\times 9\times 9$ was used.

Structural relaxation was performed with PBE using the conjugated gradients algorithm. The ionic relaxation was terminated when the change in total energy between two successive ionic steps fell below $10^{-5}$ eV.
The resulting equilibrium lattice constants of all  Heusler compounds studied here are listed in Table~\ref{tab:table1}. 

The QP$GW$ \cite{shishkin2007accurate,shishkin2007self} calculations were performed self-consistently with HSE used as the starting point.
Based on convergence tests provided in the SI, seven self-consistent $GW$ steps were performed. The total number of bands 
was set to 94 or 96, depending on the parallelization setup, approximately twice as large as the number of occupied bands for all six materials. The number of bands for which the self-energy shift was calculated 
was set to 78. The number of imaginary frequency and imaginary time grid points 
was set to 140. QP$GW$ band structures were calculated via Wannier interpolation using the \texttt{Wannier90} code \cite{pizzi2020wannier90}.  

The on-site magnetic moments were obtained by decomposing the Kohn-Sham orbitals onto the PAW projectors. 
The spin polarization percentage at the Fermi level was calculated as
$\mathrm{SP}\% = 100\,P(E_f)$, where $P(E_f)$ is given by:
\begin{equation}
    {P (E_{f})}=\frac{\int_{E_f-\Delta}^{E_f+\Delta}\left[D_{\uparrow}(E)-D_{\downarrow}(E)\right] \mathrm{d} E}{\int_{E_f-\Delta}^{E_f+\Delta}\left[D_{\uparrow}(E)+D_{\downarrow}(E)\right] \mathrm{d} E}
\end{equation}
Here, $D_{\uparrow}$ and $D_{\downarrow}$ are the spin-up (majority) and spin-down (minority) density of states (DOS). 
The integration is performed over an energy window of $2\Delta = 3.5\,kT = 0.091\,\mathrm{eV}$ centered at the Fermi level, chosen to match the full width at half maximum (FWHM) of the derivative of the Fermi-Dirac distribution at room temperature, thereby capturing the thermal broadening of the Fermi edge~\cite{wegner2006surface}. A sensitivity analysis of the SP value to the choice of integration window is provided in the SI.
A negative sign of SP\% indicates a minority spin polarization at the Fermi level. 
For the PBE, HSE, and PBE+$U$ calculations, element- and orbital-decomposed band structures were obtained by projecting the Kohn–Sham states onto PAW on-site partial-wave projectors. 
Band structure and DOS plots were generated using the open-source package \texttt{vaspvis} \cite{2022_Dardzinski_BestPracticesDFTInorganicInterfacesBO_JPHYS-CONDENSMAT}, which is available on the Python Package Index (PyPI) via \textit{pip install vaspvis}, and on GitHub at \url{https://github.com/caizefeng/vaspvis}. 
The DFT+$U$(BO) calculations were performed using the updated \texttt{BayesOpt4dftu} package, accessible at \url{https://github.com/caizefeng/BayesianOpt4dftu}.

\section{\label{sec:level3} Results and discussion}

\begin{figure*}
\centering
\includegraphics[width=1\textwidth]{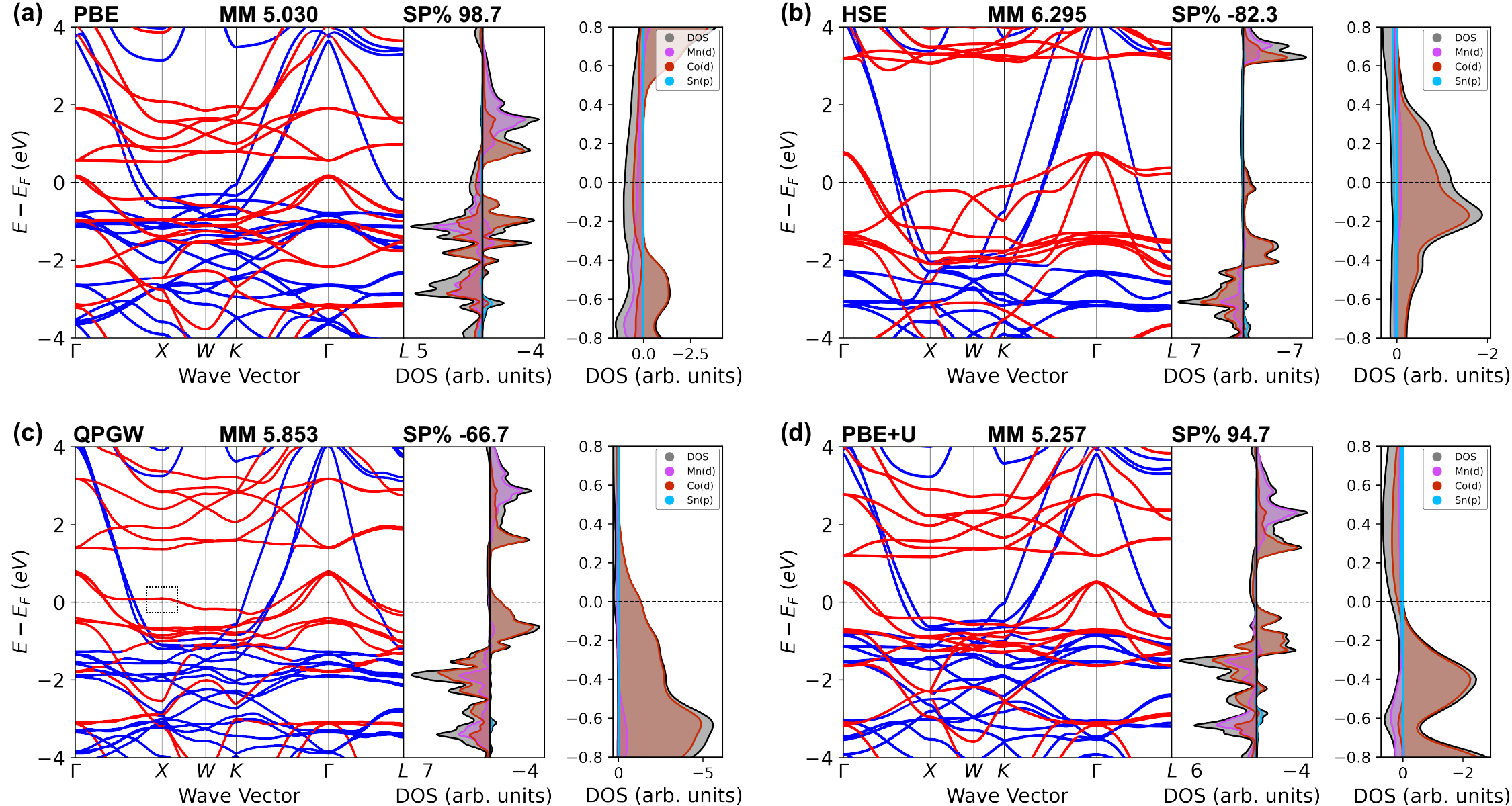}
\caption{Band structure and element-resolved DOS of Co$_2$MnSn calculated using (a) PBE, (b) HSE, (c) QP$GW$, and (d) PBE+$U$(BO). In the band-structure panels, the majority and minority spin channels are shown in blue and red, respectively. In the DOS panels, the majority and minority spin components are plotted on the left and right sides, respectively. The rightmost panels present a magnified view of the DOS around the Fermi level. The total magnetic moment (MM, in $\mu B$) and spin polarization percentage at the Fermi level (SP\%) are listed for each method. Negative spin polarization indicates that the minority spin channel is dominant around the Fermi level. 
The dashed box in (c) highlights the flat minority-spin band near the Fermi level that gives rise to the minority-dominated spin polarization in the QP$GW$ results.
}
\label{fig:band_Co2MnSn}
\end{figure*}

The Heusler alloys studied here are Co\ensuremath{_{2}}MnIn, Co\ensuremath{_{2}}MnSn, Co\ensuremath{_{2}}TiSn, Co\ensuremath{_{2}}ZrAl, Ni\ensuremath{_{2}}MnSb, and Ni\ensuremath{_{2}}MnSn. These compounds have a lattice mismatch below 1\%  with InAs or GaSb. Further screening criteria were applied to ensure that the compounds selected for this study possess a stable ferromagnetic ground state with the \textit{Fm$\bar{3}$m} symmetry at room temperature. It is worth noting that several other Co-, Ni-, or Ru-based full Heusler alloys also have nearly perfect lattice matching with InAs, but do not meet the magnetic stability criteria. For example, Co\ensuremath{_{2}}CrIn has a mismatch of only 0.02\% with InAs but has been reported to exhibit a ferrimagnetic ground state \cite{wurmehl2006co2crin}. Co\ensuremath{_{2}}NbSn, with a 0.89\% mismatch to GaSb, has a low Curie temperature of 116\,K and favors a competing \textit{Pmma} structure \cite{wolter2002structure}. Ru\ensuremath{_{2}}MnAl, Ru\ensuremath{_{2}}MnGa, and Ru\ensuremath{_{2}}MnGe have excellent lattice matching to InAs, but prefer an antiferromagnetic state at room temperature \cite{gotoh1995spin}.

Table \ref{tab:table1} presents a summary of the total magnetic moments and the degree of spin polarization around the Fermi level obtained with different electronic structure methods. According to all methods considered here, the Ni-based compounds do not have a gap in either spin channel and are not strongly spin-polarized around the Fermi level. Therefore, we focus the discussion in the main text on the Co-based compounds and provide the complete results for the Ni-based compounds in the SI. Co$_2$MnSn and Co$_2$MnIn exhibit relatively large magnetic moments due to the presence of the magnetic Mn atom in addition to Co. For these materials, the magnetic moments produced by different electronic structure methods vary significantly, with PBE giving the lowest value, HSE giving the highest value, and QP$GW$ in between.  
For Co$_2$MnSn, experimental studies report total magnetic moments in the range of 5.05-5.12~$\mu_B$ \cite{1971_Webster_HeuslerMeasureCo-basedMag_JPhysChemSolid,2000_Brown_ExpMagnetizationCo2MnXSn_JPhysACondensMatter,2015_Baral_ExpCo2MnSbPESDFT_JournAlloyComp}, which are closest to the PBE result of 5.03~$\mu_B$ and the PBE+$U$(BO) value of 5.26~$\mu_B$.
Co$_2$TiSn and Co$_2$ZrAl have smaller magnetic moments because the Y-site element is non-magnetic. For these materials the results of different electronic structure methods are more consistent. For Co$_2$TiSn, all methods produce a total magnetic moment of $\sim$2 $\mu_B$, in excellent agreement with experimental measurements~\cite{2007_Kandpal_Co2TiSnExpTheory_JournPhysDApplPhys,2015_Bainsla_Co2TiSnMagMonSP_CurrentApplPhys,2010_Barth_Co2TiSnMagMomTheoryPES_PRB,2017_Ooka_MagSpinPolCo2TiSn_IEEEMagLet,1972_Webster_MagMomCo2TiX=Sn_JournChemSolid}. 
For Co$_2$ZrAl, all methods (except HSE) predict a magnetic moment close to 1 $\mu_B$, which is slightly higher than the experimentally measured value of approximately 0.7 $\mu_B$~\cite{2005_Kanomata_MagMomCo2ZrAl_KournAlloyCompo,1981_Buschow_HeuslerCo2ZrAlMagMomLat_JournMagMat,1973_Ziebeck_Co2Zr-NbMagMonLatCon_JournPhysChem}. 
The measured magnetic moments of Ni$_2$MnSb range from 3.22 to 4.31~$\mu_B$ \cite{1970_Shinohara_Ni2MnX,1983_Buschow_LotsMaterialsNi2MnX}. The scatter may be due to differences in sample quality or stoichiometry. The computed values range from 4.04~$\mu_B$ with PBE to 4.66~$\mu_B$ with QP$GW$.  
Overall, both HSE and QP$GW$ tend to overestimate the total magnetic moments relative to experiment for several compounds.
Of the four Co-based compounds, only Co$_2$TiSn is unanimously predicted by all methods to be a half-metal. In the following, we present a detailed analysis of the origin of the differences between electronic structure methods.  

Figure~\ref{fig:band_Co2MnSn} shows the band structure and element-resolved DOS of Co$_2$MnSn obtained with PBE, HSE, QP$GW$, and PBE+$U$(BO).
With the PBE functional, the DOS around the Fermi level is mainly contributed by the Co $3d$ and Mn $3d$ states. The DOS shows a half-metallic gap of 0.406 eV in the minority-spin channel (on the right-hand side of the DOS panels) and located 0.162 eV above the Fermi level. 
It should be noted that the position and width of the gap are determined from the band structure rather than from the DOS. Although the minority-spin gap appears sizable and may seem to intersect the Fermi level in the DOS plots of Figure~\ref{fig:band_Co2MnSn}(a), 
a small but nonzero minority-spin density remains at the Fermi level, leading to a spin polarization below 100\% (98.7\% in this case). The apparent discrepancy arises because the band structure is plotted along selected high-symmetry paths, which do not necessarily represent the full Brillouin-zone integration that determines the total DOS.

The HSE functional yields a considerably different band structure than PBE, as shown in panel b. The Mn $d$ states are strongly exchange-split and shifted farther in energy above and below the Fermi level. The gap of 2.40 eV in the minority spin channel is considerably wider than with PBE and it is located 0.509 eV above the Fermi level. The DOS around the Fermi level is contributed predominantly by the Co $3d$ states, leading to minority spin polarization of -82.3\%. The behavior of HSE likely originates from the overestimation of exchange splitting in $d$-electron systems by the non-local exchange term, which is a known limitation of this functional~\cite{paier2006screened, paier2007does}.

With QP$GW$ (panel c), the exchange splitting of the Mn $d$ states is not as large as with HSE. The gap in the minority spin channel is 0.818 eV and it is located 0.476 eV above the Fermi level.  The DOS around the Fermi level is dominated by the Co $d$ states, similar to HSE, with a small contribution from the Mn $d$ states.  This results in a minority spin polarization of -66.7\%. Although the QP$GW$ band structure appears overall more similar to PBE than to HSE, a crucial difference is the rather flat minority band that lies slightly above the Fermi level around the X-point (indicated by the box in panel c). This is sufficient to flip the spin polarization around the Fermi level. 

The results of PBE+$U$(BO) are in between PBE and QP$GW$, as shown in panel d. The gap in the minority spin channel is 0.905 eV, similar to QP$GW$, but its lower edge is located right around the Fermi level. The DOS around the Fermi level is dominated by the Co $d$ manifold. However, the flat minority band around the X-point lies below the Fermi level, leading to a majority spin polarization of 94.7\%, similar to PBE. The interplay between the relative positions of the Co and Mn states $d$ can thus lead to qualitatively different results with different electronic structure methods. A spin polarization of about 60\% has been measured for Co$_2$MnSn by point-contact Andreev reflection (PCAR) ~\cite{2009_Varaprasad_ActaMater}. We note that PCAR measures only the absolute value of the spin polarization irrespective of its sign~\cite{soulen1998measuring,nadgorny2003measurements}. The absolute value of the spin polarization obtained from QP$GW$ is close to the PCAR experiment, but we cannot determine whether the minority spin polarity is correct.   This leaves us with the questions of whether Co$_2$MnSn is a half-metal, what is the dominant spin channel at the Fermi level, and which method should we trust?

\begin{figure}
\centering
\includegraphics[width=1\columnwidth]{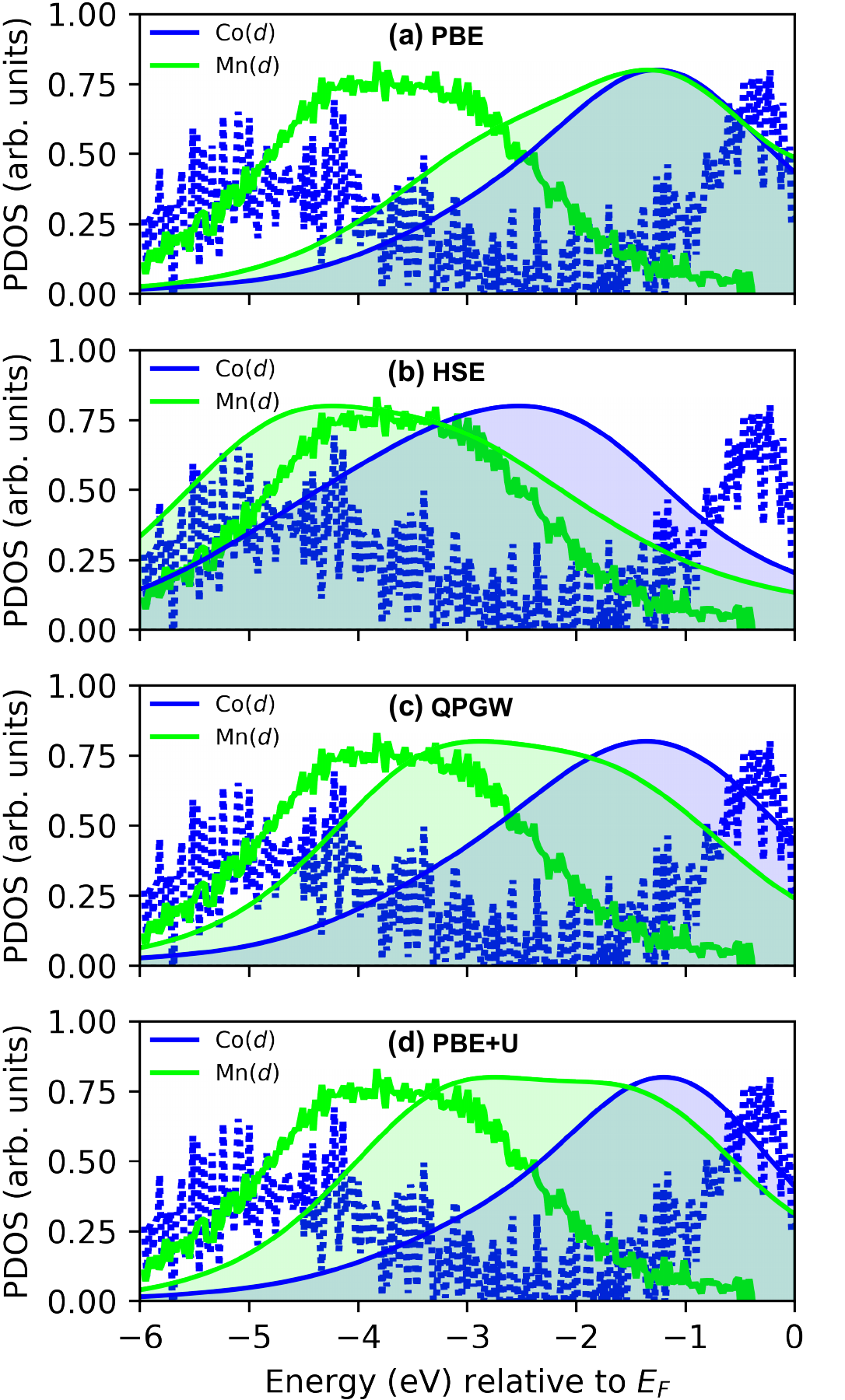}
\caption{Orbital-resolved DOS of Co$_2$MnSn calculated using (a) PBE, (b) HSE, (c) QP$GW$, and (d) PBE+$U$(BO), compared to the resonant photoemission spectroscopy (RPES) spectra of Co-3$d$ and Mn-3$d$, reproduced with permission from Ref. \cite{2015_Baral_ExpCo2MnSbPESDFT_JournAlloyComp}. The computed spectra are broadened by a Gaussian with $\sigma = 0.77\,\mathrm{eV}$ to simulate the resolution of the experiment. All spectra are normalized to the same maximum intensity. The satellite features between $-6\,\mathrm{eV}$ and $-2\,\mathrm{eV}$ in the experimental spectrum of Co-3$d$ are attributed in Ref. \cite{2015_Baral_ExpCo2MnSbPESDFT_JournAlloyComp} to excitation effects that are not captured by the electronic structure methods used here.
}
\label{fig:dos_Co2MnSn}
\end{figure}

To attempt to answer these questions, the orbital-resolved DOS of Co$_2$MnSn is compared with resonant photoemission spectroscopy (RPES) data~\cite{2015_Baral_ExpCo2MnSbPESDFT_JournAlloyComp} in Figure~\ref{fig:dos_Co2MnSn}. The experimental data shown are difference spectra that were obtained therein by subtracting the off-resonance spectrum from the on-resonance spectrum to resolve the partial DOS of the Co-3$d$ and Mn-3$d$ states. The RPES data shows that the DOS close to the Fermi level is predominantly contributed by the Co-3$d$, whereas the Mn-3$d$ states lie lower in energy, around -3.8 eV below the Fermi level. The satellite feature between -6 eV and -2 eV in the Co-3$d$ RPES spectrum was attributed in Ref.~\cite{2015_Baral_ExpCo2MnSbPESDFT_JournAlloyComp} to excited-state effects that are not captured by the electronic structure methods used here. We focus the comparison to experiment on the Co-3$d$ and Mn-3$d$ main peak positions. With PBE the Co-3$d$ and Mn-3$d$ overlap, which clearly contradicts the experimental data.  HSE shifts the Mn-3$d$ peak to around -4 eV, which is close to the experiment, but the Co-3$d$ states are shifted too deep to around -2.6 eV below the Fermi level, which deviates significantly from the experiment. With QP$GW$ the Mn-3$d$ peak is slightly too high and the Co-3$d$ peak is slightly too low, providing the closest match to experiment of the methods considered here. The PBE+$U$(BO) partial DOS is very similar to QP$GW$. In Ref.~\cite{2015_Baral_ExpCo2MnSbPESDFT_JournAlloyComp} it was found that applying a Hubbard $U$ correction of 3 eV to the Mn-3$d$ states and no correction to the Co-3$d$ provided the closest agreement with experiment. This is different than the $U$ values of 1.512 eV for Co-3$d$ and 1.652 eV for Mn-3$d$, which best reproduce the QP$GW$ results.  
Detailed experimental characterization with spin-resolved angle-resolved photoemission spectroscopy (ARPES) \cite{zhang2022angle, iwasawa2024efficiency} would be needed to unambiguously determine which electronic structure method is the most reliable.

\begin{figure*}
\centering
\includegraphics[width=1\textwidth]{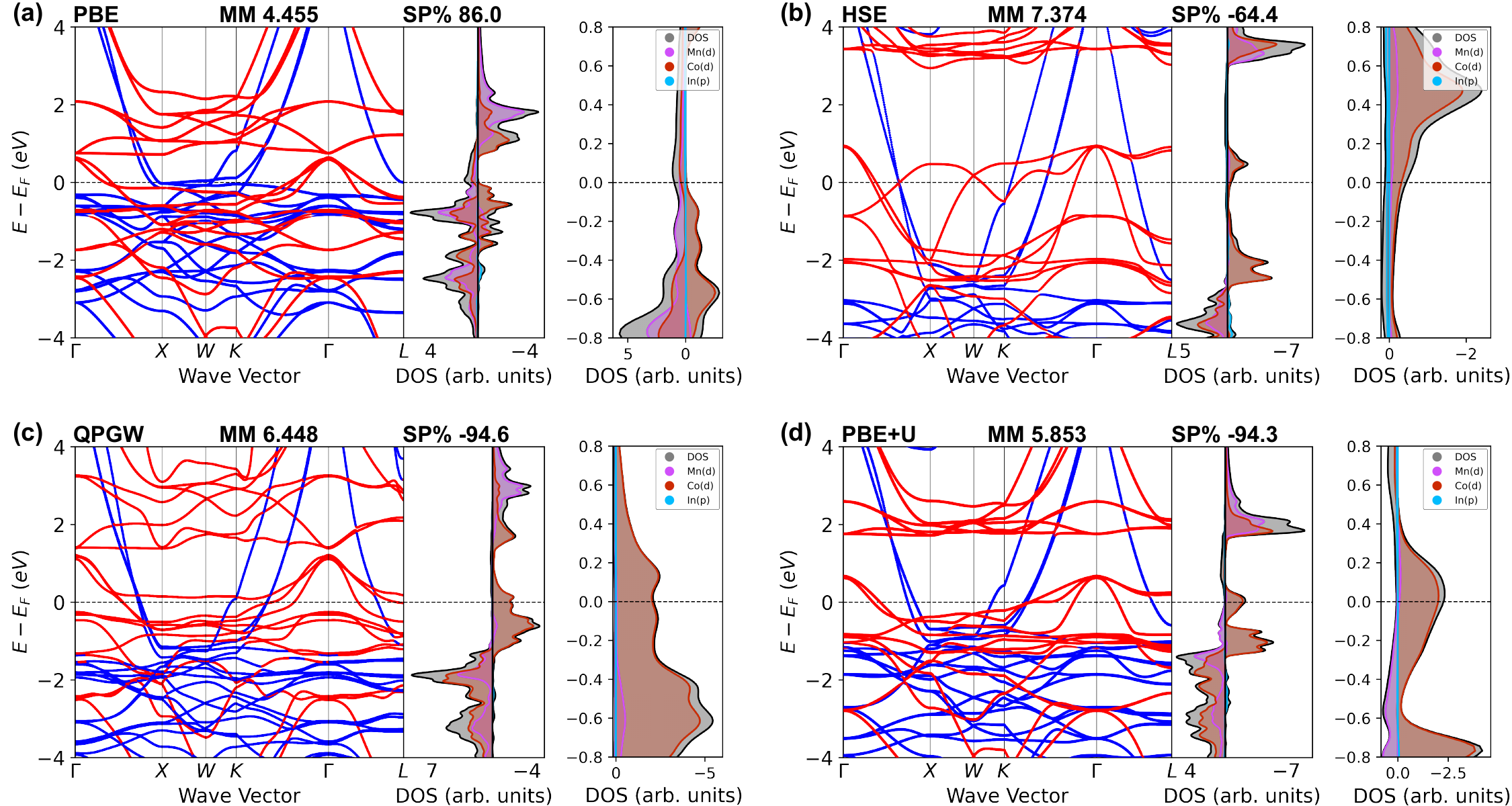}
\caption{Band structure and element-resolved DOS of Co$_2$MnIn calculated using (a) PBE, (b) HSE, (c) QP$GW$, and (d) PBE+$U$(BO). In the band-structure panels, the majority and minority spin channels are shown in blue and red, respectively. In the DOS panels, the majority and minority spin components are plotted on the left and right sides, respectively. The rightmost panels present a magnified view of the DOS around the Fermi level. The total magnetic moment (MM, in $\mu B$) and spin polarization percentage at the Fermi level (SP\%) are listed for each method. Negative spin polarization indicates that the minority spin channel is dominant around the Fermi level.}
\label{fig:band_Co2MnIn}
\end{figure*}

The case of Co$_2$MnIn, shown in Figure~\ref{fig:band_Co2MnIn}, is similar to Co$_2$MnSn in the sense that the interplay between the positions of the Co-3$d$ and Mn-3$d$ states leads to inconsistent results from different electronic structure methods. 
With PBE there is a very small gap of 0.103 eV in the minority-spin channel. The lower edge of this gap lies approximately 0.627~eV above the Fermi level. Similar to Co$_2$MnSn, there are contributions from both Co and Mn $d$ states near the Fermi level, resulting in a spin polarization of 86.0\%. Here, too, the position and width of the minority-spin gap are determined from the band structure rather than from the DOS panels of Figure~\ref{fig:band_Co2MnIn}.

With HSE, the Mn $d$ states are strongly exchange-split and pushed away from the Fermi level. The gap of 1.97 eV in the minority spin channel is much larger than with PBE and found 0.997 eV above the Fermi level. The main peaks of the Co $d$ states are also pushed away from the Fermi level, leaving a relatively small minority spin density tail. The In $p$ states contribute to the majority channel at the Fermi level, resulting in a minority spin polarization of -64.4\%.

With QP$GW$ the gap in the minority spin channel is 0.196 eV and it is located 1.199 eV above the Fermi level. The Mn $d$ states are not near the Fermi level, although not as far away as with HSE. The DOS at the Fermi level is dominated by a large peak of minority spin Co $d$ states, yielding a high spin polarization of -94.6\%. 

PBE+$U$(BO) produces a gap of 0.669 eV in the minority spin channel, which lies 0.499 eV above the Fermi level. The relative positions of the Co 3$d$ and Mn 3$d$ states is similar to QP$GW$ with the DOS around the Fermi level dominated by a large Co 3$d$ minority spin peak. Unlike pure PBE, which favors majority-spin polarization, PBE+$U$(BO) yields the same minority-spin polarization of -94.3\% as QP$GW$.
If we consider the results of QP$GW$ to be the most reliable, then although Co$_2$MnIn is not a half-metal, 
it has a high spin polarization and a high density of states at the Fermi level, which could be advantageous for spintronic applications.

\begin{figure*}
\centering
\includegraphics[width=1\textwidth]{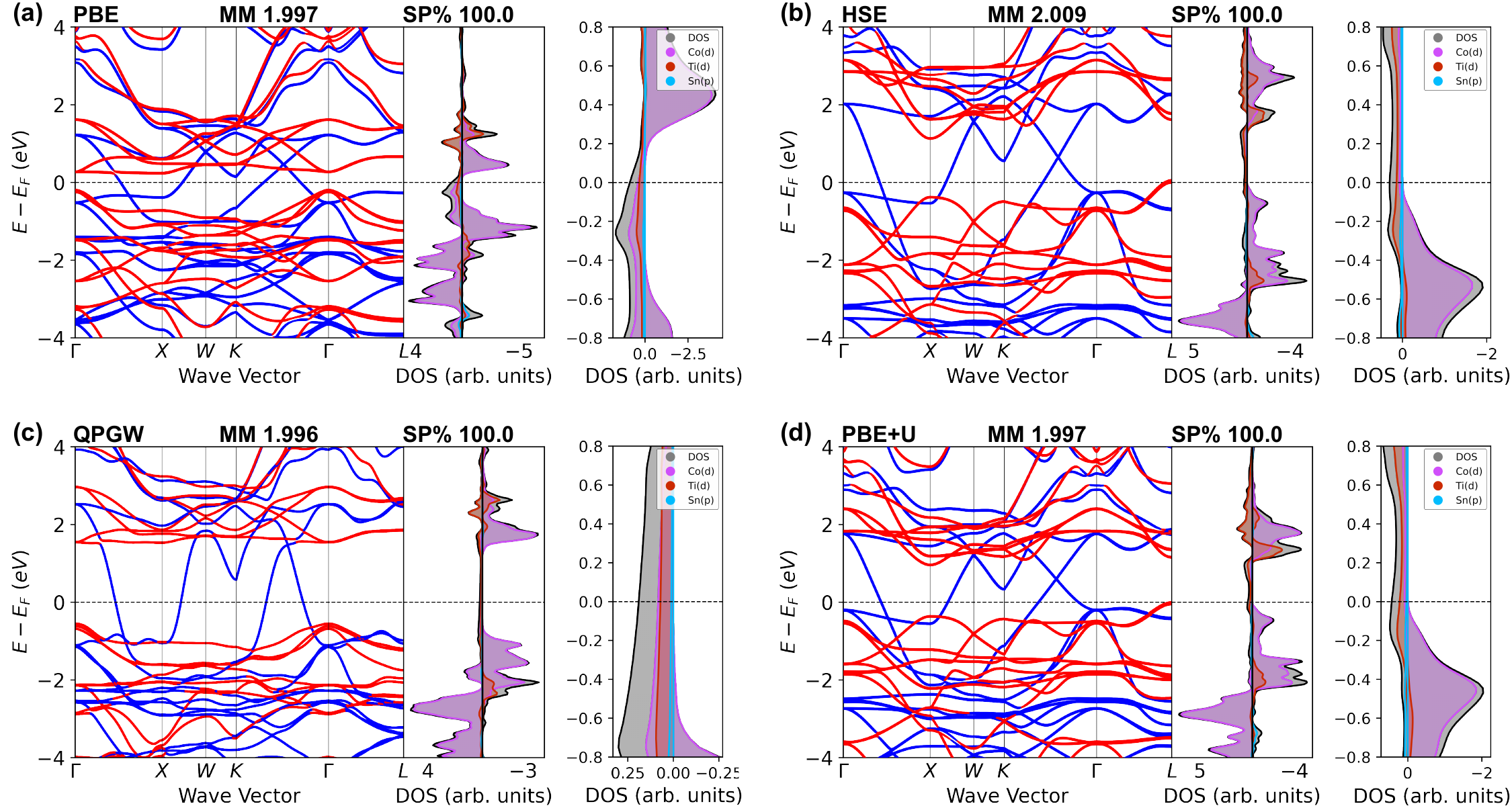}
\caption{Band structure and element-resolved DOS of Co$_2$TiSn calculated using (a) PBE, (b) HSE, (c) QP$GW$, and (d) PBE+$U$(BO). In the band-structure panels, the majority and minority spin channels are shown in blue and red, respectively. In the DOS panels, the majority and minority spin components are plotted on the left and right sides, respectively. The rightmost panels present a magnified view of the DOS around the Fermi level. The total magnetic moment (MM, in $\mu B$) and spin polarization percentage at the Fermi level (SP\%) are listed for each method.}
\label{fig:band_Co2TiSn}
\end{figure*}

As mentioned above, Co$_2$TiSn, which has long been regarded as a prototypical example of a half-metallic Heusler compound ~\cite{miura2006ab, 2005_Lee_Co2TiX, 2006_Hickey_Co2TiSnGGATheoryFermiSurface_JournPhys3ACondensMatter}, is the only material unanimously predicted by all methods used here to be a half-metal. As shown in Figure~\ref{fig:band_Co2TiSn}, there are still significant differences between the band structures produced by different methods with respect to the size of the gap in the minority spin channel and its position in energy and in $k$-space. PBE yields a gap of 0.605 eV, which is significantly smaller than the values of 1.291 eV, 1.563 eV, and 1.016 eV produced by HSE, QP$GW$, and PBE+$U$(BO), respectively. In addition, with PBE and QP$GW$ the gap is located at the $\Gamma$-point, whereas with HSE and PBE+$U$(BO) it is located at the L-point. These differences in the band structure do not affect the outcome that only the majority spin Co $d$ states contribute to the DOS at the Fermi level, which results in a 100\% spin polarization.
Experimentally, however, spin polarization values obtained via point-contact Andreev reflection (PCAR) are consistently lower, with reported values of 64$\pm$2\%, 57\%, and 58.9$\pm$2.2\%~\cite{2015_Bainsla_Co2TiSnMagMonSP_CurrentApplPhys,Varaprasad_2012_Co2FeGaCo2TiSnSPVal_ActaMater,2017_Ooka_MagSpinPolCo2TiSn_IEEEMagLet}. A possible explanation is that the spin polarization measured by PCAR is highly sensitive to perturbations such as defects, disorder, surface effects, or deviations from stoichiometry~\cite{klaer2011element}, even when the minority-spin gap edge is not located particularly close to the Fermi level.

\begin{figure*}
\centering
\includegraphics[width=1\textwidth]{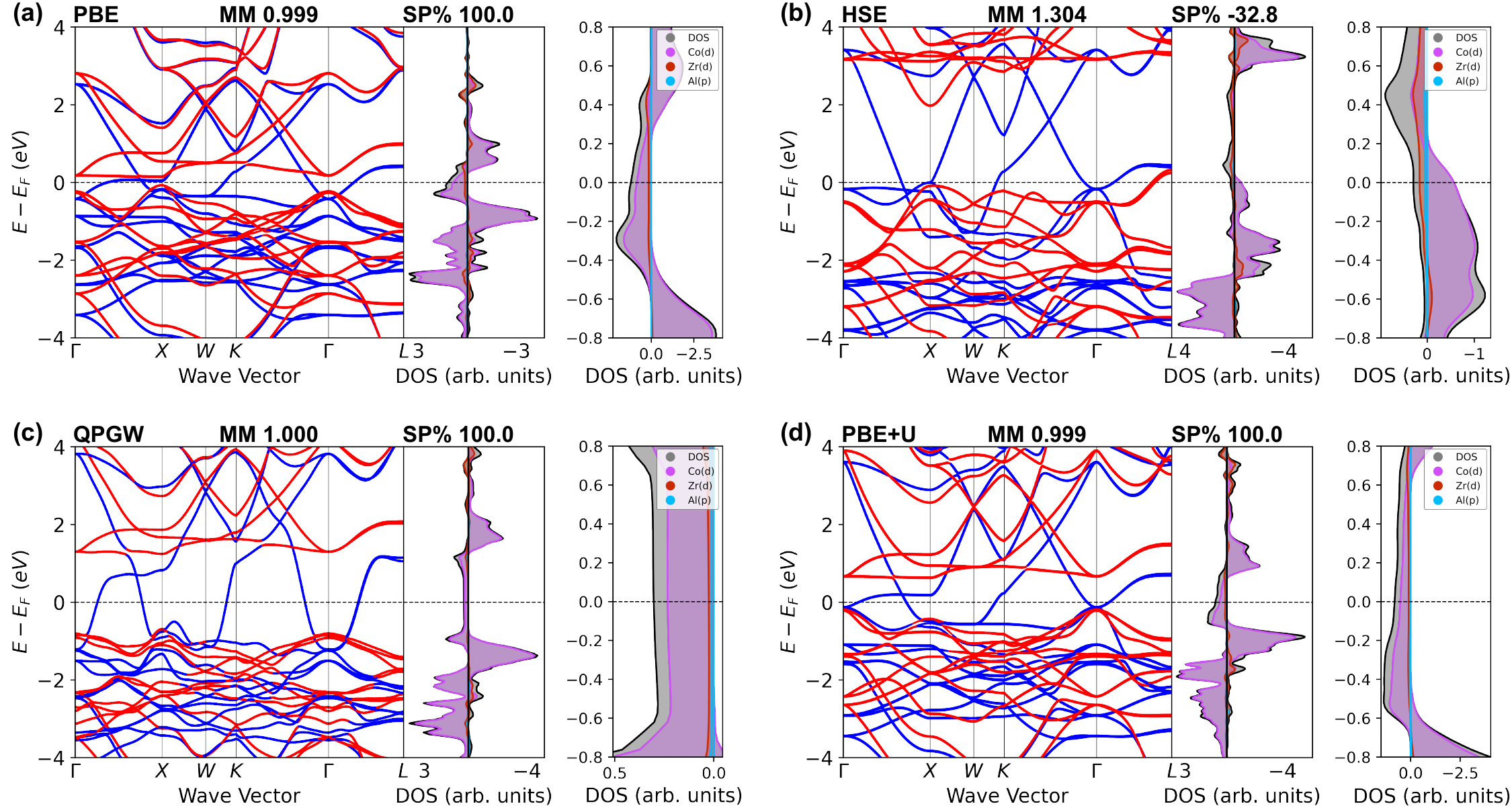}
\caption{Band structure and element-resolved DOS of Co$_2$ZrAl calculated using (a) PBE, (b) HSE, (c) QP$GW$, and (d) PBE+$U$(BO). In the band-structure panels, the majority and minority spin channels are shown in blue and red, respectively. In the DOS panels, the majority and minority spin components are plotted on the left and right sides, respectively. The rightmost panels present a magnified view of the DOS around the Fermi level. The total magnetic moment (MM, in $\mu B$) and spin polarization percentage at the Fermi level (SP\%) are listed for each method. Negative spin polarization indicates that the minority spin channel is dominant around the Fermi level.}
\label{fig:band_Co2ZrAl}
\end{figure*}

Co$_2$ZrAl is predicted to be a half-metal by PBE, QP$GW$, and PBE+$U$(BO). With all three methods, only the majority spin Co $d$ states contribute to the DOS at the Fermi level, leading to a 100\% spin polarization, as shown in Figure \ref{fig:band_Co2ZrAl}. 
However, these three methods differ in the details of the band structure, particularly in the magnitude and location in $k$-space  of the minority-spin gap. With PBE, the minority-spin gap is 0.515~eV, whereas QP$GW$ predicts a substantially larger gap of 1.913~eV.  The PBE+$U$(BO) result of 1.044~eV lies between these two values. In addition, with PBE and PBE+$U$(BO) the minority-spin gap is indirect between the X and $\Gamma$ points, whereas with QP$GW$ it is direct at the X-point.
HSE produces a markedly different picture. The majority spin Co $d$ states are pushed away from the Fermi level. As a result, the DOS at the Fermi level is dominated by minority-spin Co $d$ states, with additional contributions from majority-spin Zr $d$ states and only a marginal contribution of Al $p$ states, leading to an overall minority spin polarization of -32.8\%.

Finally, we comment on the practical utility of the DFT+$U$(BO) approach. The QP$GW$ reference is a one-time cost, incurred once for the bulk primitive cell. The resulting $U$ can then be transferred to strained structures, surfaces, interfaces, and large supercells, for which a direct QP$GW$ (or even hybrid functional) treatment is not feasible. In the SI, we further compare DFT+$U$(BO) with the linear-response approach to determining $U$ \cite{2005_Cococcioni_LinearResponseU_PRB}. For the six Heuslers studied here,  converging the linear-response procedure is more computationally expensive than the entire BO workflow including its QP$GW$ reference. Moreover, the linear response approach tends to yield large positive $U$ values, which do not reproduce the QP$GW$ band structures and magnetic moments as closely as  $U$(BO). Our results are consistent with the previously reported tendency of the linear-response approach to overestimate $U$ for the delocalized Co $d$ states in Co-based Heuslers \cite{2019_Nawa_DFTUCoBasedHeusler_RSCAdv}.

\section{\label{sec:level4} Conclusion}
In summary, we investigated the electronic structure and magnetic properties of six Heusler compounds in the cubic $L2_1$ (\textit{Fm$\bar{3}$m}) phase with a stable ferromagnetic ground state that are lattice-matched to InAs and GaSb. We compared the results obtained using PBE, HSE, QP$GW$, and PBE+$U$(BO) with $U$ values optimized to reproduce as closely as possible the QP$GW$ band structure and magnetic moments. The results reveal a strong dependence of the magnitude and position of the gap in the minority spin channel, as well as the spin polarization at the Fermi level, on the electronic structure method used. In some cases, the predictions of different methods even differ qualitatively with respect to whether a material is expected to be a half-metal and the dominant spin channel (majority or minority) at the Fermi level. Our analysis shows that these differences arise from the treatment of $d$ states with some systematic trends among methods. PBE tends to underestimate, whereas HSE tends to overestimate, the bandwidths and exchange splitting of the $d$ manifolds, particularly in materials with strong $d$–$d$ hybridization. The results of QP$GW$  tend to be in the range between PBE and HSE. In most cases, PBE+$U$(BO) reproduces well the salient features of the QP$GW$ band structure. The only exceptions found here are Co$_2$MnSn and Ni$_2$MnSb, for which the spin polarization at the Fermi level obtained with PBE+$U$(BO) differs qualitatively from QP$GW$. This may occur because the spin polarization at the Fermi level is not included in the BO objective function.  

Of the materials considered here, we expect Co$_2$TiSn and Co$_2$ZrAl to be half-metals with a high degree of confidence, based on consistent predictions of all or most of the methods used here.
Our findings have significant implications for the reliability of high-throughput studies of magnetic Heuslers and related materials based on semi-local DFT. We recommend further screening of candidate materials using more accurate methods prior to pursuing them experimentally. For simulations of interfaces and other large systems involving Heuslers, we suggest PBE+$U$(BO) as a practical solution for obtaining qualitatively correct results at a reasonable computational cost. 
This study also highlights the critical need for probing the band structure of Heuslers and related materials by ARPES, preferably spin resolved, in order to assess the accuracy and reliability of electronic structure methods for this important class of materials.

\begin{acknowledgments}
We thank Paul Crowell from the University of Minnesota for helpful discussions. This research was funded by the Department of Energy through grant DE-SC0019274. This research used resources of the National Energy Research Scientific Computing Center (NERSC), a DOE Office of Science User Facility supported by the Office of Science of the U.S. Department of Energy under contract no. DE-AC02-05CH11231.
\end{acknowledgments}

\par\vspace{16pt}
\centerline{\normalfont\bfseries\small DATA AVAILABILITY}
\vspace{8pt}\par\noindent
The VASP input and output files of the calculations presented here are available through Zenodo at \url{zenodo.org/records/18705349} with DOI 10.5281/zenodo.18705349.
 The updated \texttt{BayesOpt4dftu} package is available at \url{github.com/caizefeng/BayesianOpt4dftu}.
 The \texttt{vaspvis} package is available on the Python Package Index (PyPI) via \textit{pip install vaspvis}, and on GitHub at \url{github.com/caizefeng/vaspvis}.

\bibliography{reference}

\end{document}


\title{\Large \textbf{Supplemental Information: Combining Quasiparticle Self-Consistent $GW$ and Machine-Learned DFT+$U$ to Assess Half-Metallicity in Co- and Ni-Based Heuslers}}

\author{Zefeng Cai}
\affiliation{Department of Materials Science and Engineering, Carnegie Mellon University, Pittsburgh, PA 15213, USA}
\author{Malcolm J. A. Jardine}
\affiliation{Department of Materials Science and Engineering, Carnegie Mellon University, Pittsburgh, PA 15213, USA}
\author{Maituo Yu}
\affiliation{Department of Materials Science and Engineering, Carnegie Mellon University, Pittsburgh, PA 15213, USA}
\author{Chenbo Min}
\affiliation{Department of Materials Science and Engineering, Carnegie Mellon University, Pittsburgh, PA 15213, USA}
\author{Jiatian Wu}
\affiliation{Department of Materials Science and Engineering, Carnegie Mellon University, Pittsburgh, PA 15213, USA}
\author{Hantian Liu}
\affiliation{Department of Materials Science and Engineering, Carnegie Mellon University, Pittsburgh, PA 15213, USA}
\author{Derek Dardzinski}
\affiliation{Department of Materials Science and Engineering, Carnegie Mellon University, Pittsburgh, PA 15213, USA}
\author{Christopher J. Palmstr{\o}m}
\affiliation{Materials Department, University of California-Santa Barbara, Santa Barbara, CA 93106, USA}
\affiliation{Department of Electrical and Computer Engineering, University of California-Santa Barbara, Santa Barbara, CA 93106, USA}
\author{Noa Marom}
 \email{Electronic mail: nmarom@andrew.cmu.edu}
\affiliation{Department of Materials Science and Engineering, Carnegie Mellon University, Pittsburgh, PA 15213, USA}
\affiliation{Department of Chemistry, Carnegie Mellon University, Pittsburgh, PA 15213, USA}
\affiliation{Department of Physics, Carnegie Mellon University, Pittsburgh, PA 15213, USA}

\date{\today}

\maketitle

\tableofcontents

\newpage
\section{\label{sec:level1} Results for Additional Materials}

\begin{figure*}[h]
\centering
\includegraphics[width=1\textwidth]{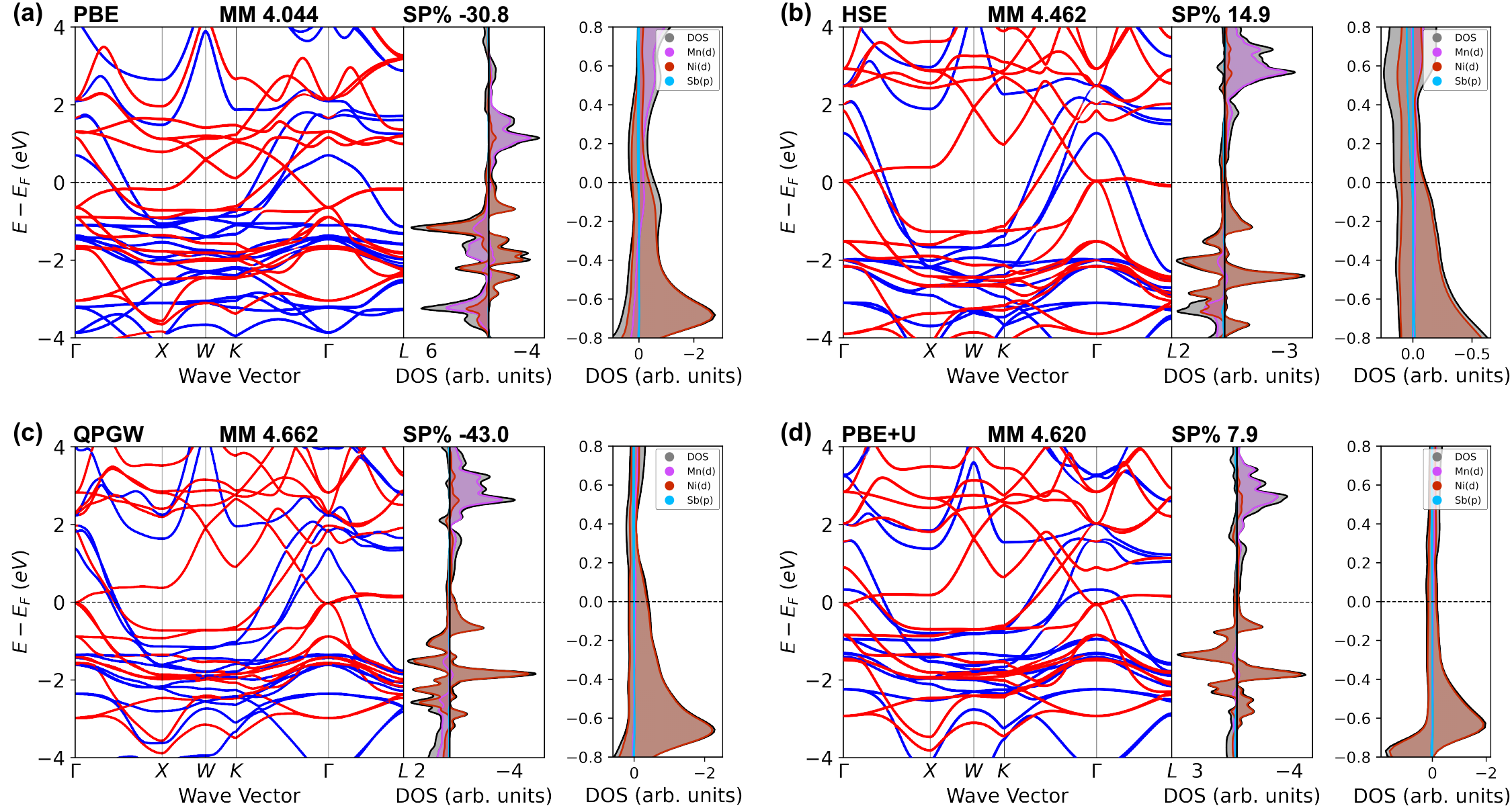}
\caption{Band structure and orbital-decomposed density of states (DOS) of Ni$_2$MnSb calculated using (a) PBE, (b) HSE, (c) QP$GW$, and (d) PBE+$U$(BO). In the band structure panels, majority and minority spin channels are shown in blue and red, respectively. The right panels show a magnified view of the DOS around the Fermi level. The total magnetic moment (MM, in $\mu B$) and spin polarization percentage at the Fermi level (SP\%) are listed for each method. Negative spin polarization indicates that the minority spin channel is dominant around the Fermi level.}
\end{figure*}

The Ni-based compounds, Ni$_2$MnSb and Ni$_2$MnSn, are not expected to be half-metals according to all methods because no gap is found in the minority spin channel. However, there are still notable differences in the electronic structure produced by different methods, in particular with respect to the relative position of the Ni and Mn $d$ states. Similar to the trends found for the Co-based compounds, with PBE there are contributions from both the Ni 3$d$ and the Mn 3$d$ orbitals around the Fermi level. Contributions from both the Ni and Mn 3$d$ minority DOS lead to a minority spin polarization at the Fermi level.  HSE pushes the Mn 3$d$ states away from the Fermi level, leaving only the Ni 3$d$ states around the Fermi level. For Ni$_2$MnSb, the main peak of the Ni 3$d$ minority channel is also pushed away from the Fermi level, which leads to a switch of the spin polarization from minority to majority. For Ni$_2$MnSn, the changes in the band structure do not lead to a significant change in the spin polarization around the Fermi level. With QP$GW$ the Mn 3$d$ states are not as far from the Fermi level as with HSE and the Ni 3$d$ minority peak remains close to the Fermi level, leading to a minority spin polarization for both materials. For Ni$_2$MnSb, although the PBE+$U$(BO) band structure resembles the QP$GW$ result, the small majority spin polarization at the Fermi level is closer to the HSE value. As noted in the main text, this may occur because the spin polarization at the Fermi level is not included in the BO objective function, only the total magnetic moment. For Ni$_2$MnSn, PBE+$U$(BO) reproduces well the QP$GW$ band structure and the spin polarization at the Fermi level.  

\vspace{2em}

\begin{figure*}[h]
\centering
\includegraphics[width=1\textwidth]{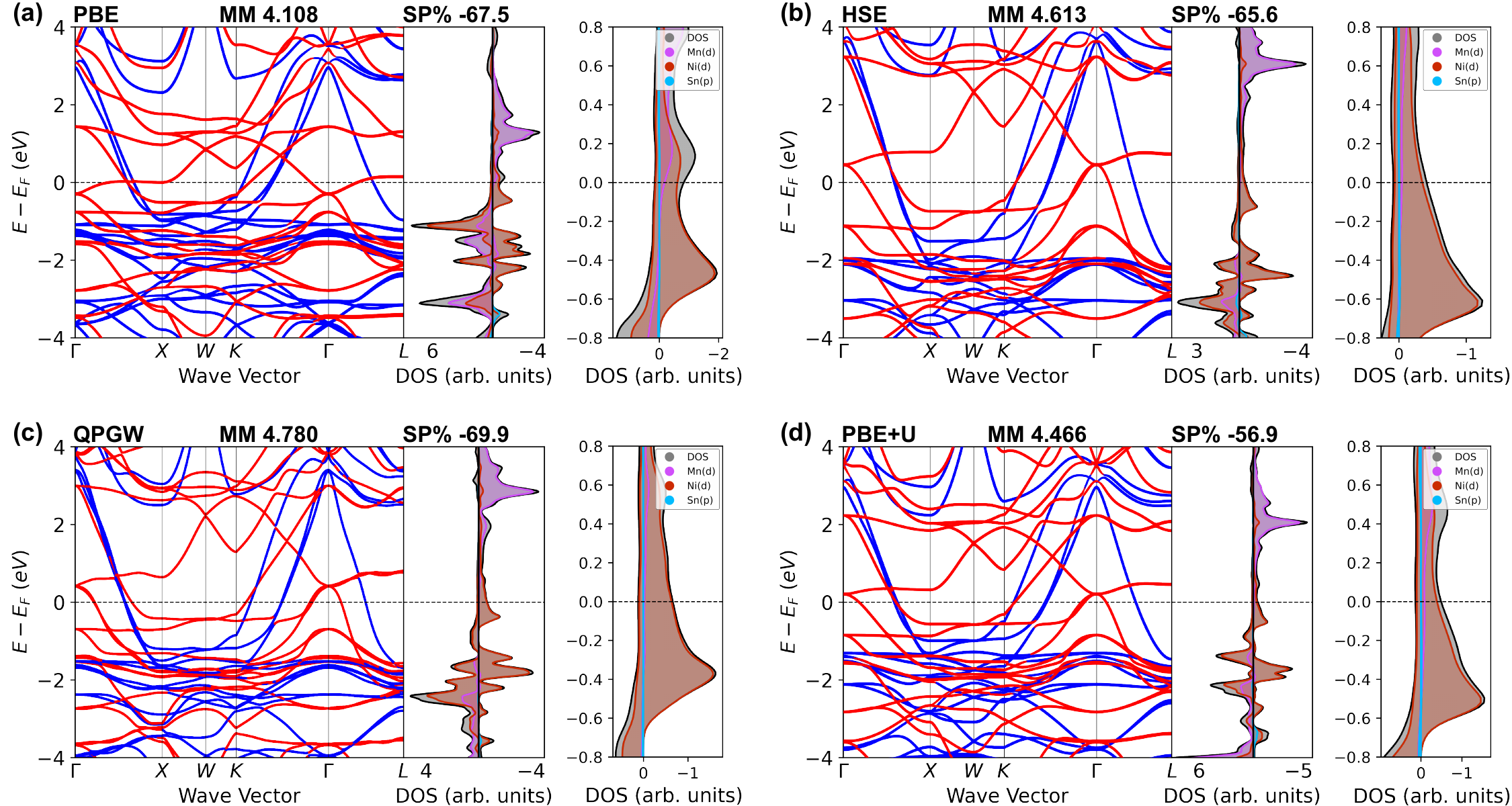}
\caption{Band structure and orbital-decomposed density of states (DOS) of Ni$_2$MnSn calculated using (a) PBE, (b) HSE, (c) QP$GW$, and (d) PBE+$U$(BO). In the band structure panels, majority and minority spin channels are shown in blue and red, respectively. The right panels show a magnified view of the DOS around the Fermi level. The total magnetic moment (MM, in $\mu B$) and spin polarization percentage at the Fermi level (SP\%) are listed for each method. Negative spin polarization indicates that the minority spin channel is dominant around the Fermi level.}
\end{figure*}

\clearpage
\section{Comparison with the Linear-Response Hubbard $U$}\label{sec:LR}

To place the BO-determined $U$ values in context, we also determined the Hubbard $U$ for all six compounds using the linear-response (LR) approach of Cococcioni and de Gironcoli \cite{2005_Cococcioni_LinearResponseU_PRB}. A rigid potential shift $\alpha$ was applied to the $d$ shell of one symmetry-inequivalent transition-metal site (X or Y), split off as a separate species, in a $3\times3\times3$ (108-atom) supercell. The response of the $d$-orbital occupation to $\alpha = \pm 0.05, \pm 0.10, \pm 0.15, \pm 0.20$~eV was fitted linearly, with the bare response $\chi_0$ obtained from non-self-consistent calculations at frozen charge density and the fully screened response $\chi$ from self-consistent calculations, and $U = \chi^{-1} - \chi_0^{-1}$. All LR calculations were noncollinear and included SOC, using a $\Gamma$-centered $3\times3\times3$ $k$-mesh (the same $k$-point spacing of 0.03~$2\pi$/\AA{} as the other calculations in this work), Methfessel--Paxton smearing of 0.15~eV, and the same PAW potentials and plane-wave cutoff as the other calculations in this work. The $2\times2\times2$ (32-atom) supercell yields $U$ values within 0.3~eV of the $3\times3\times3$ results for all sites, indicating adequate size convergence of the values reported here.

\begin{figure}[H]
\def\baselinestretch{1}\footnotesize
\centering
\begin{minipage}[t]{0.31\textwidth}\centering
\includegraphics[width=\linewidth]{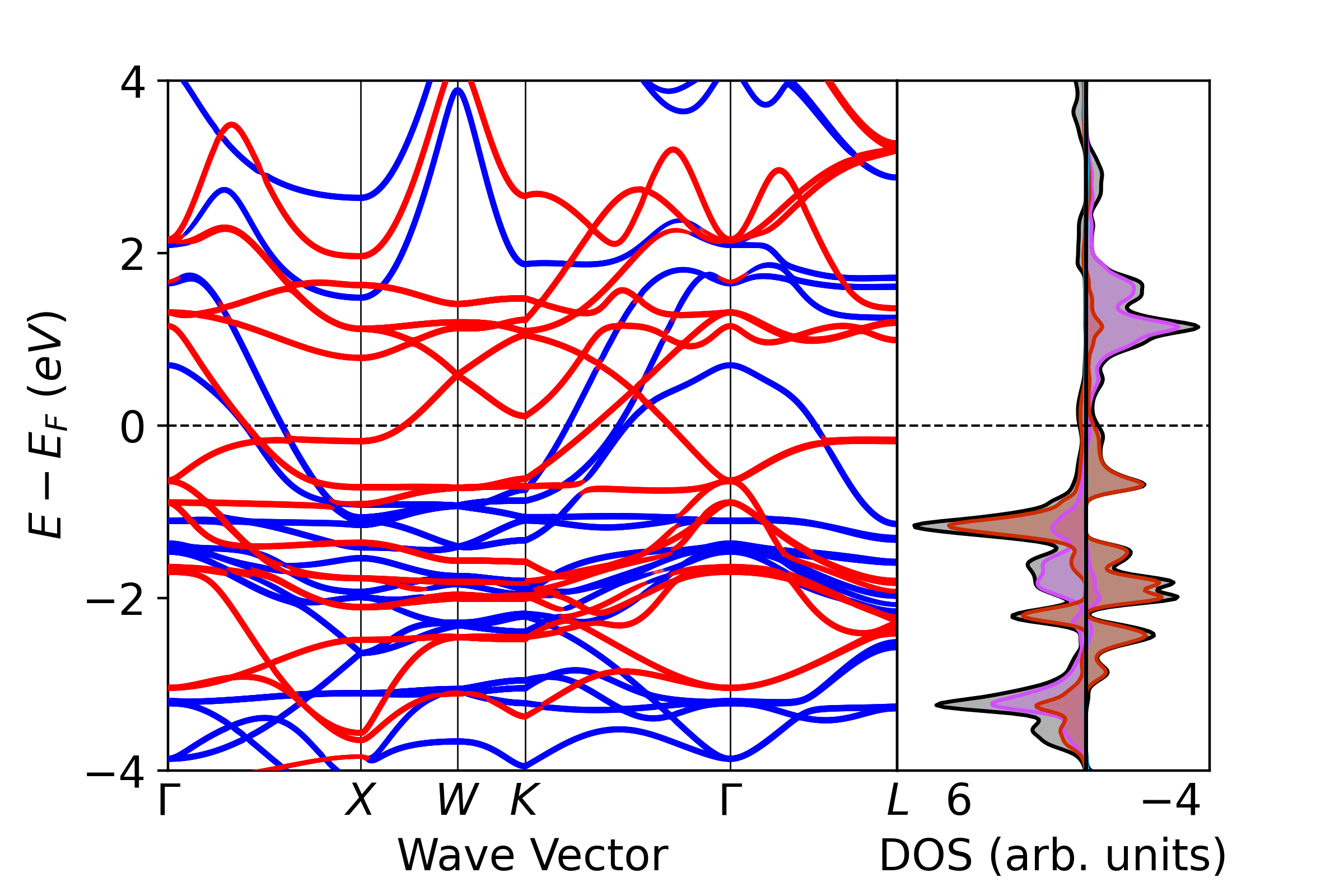}\\ (a) PBE
\end{minipage}\hspace{1em}
\begin{minipage}[t]{0.31\textwidth}\centering
\includegraphics[width=\linewidth]{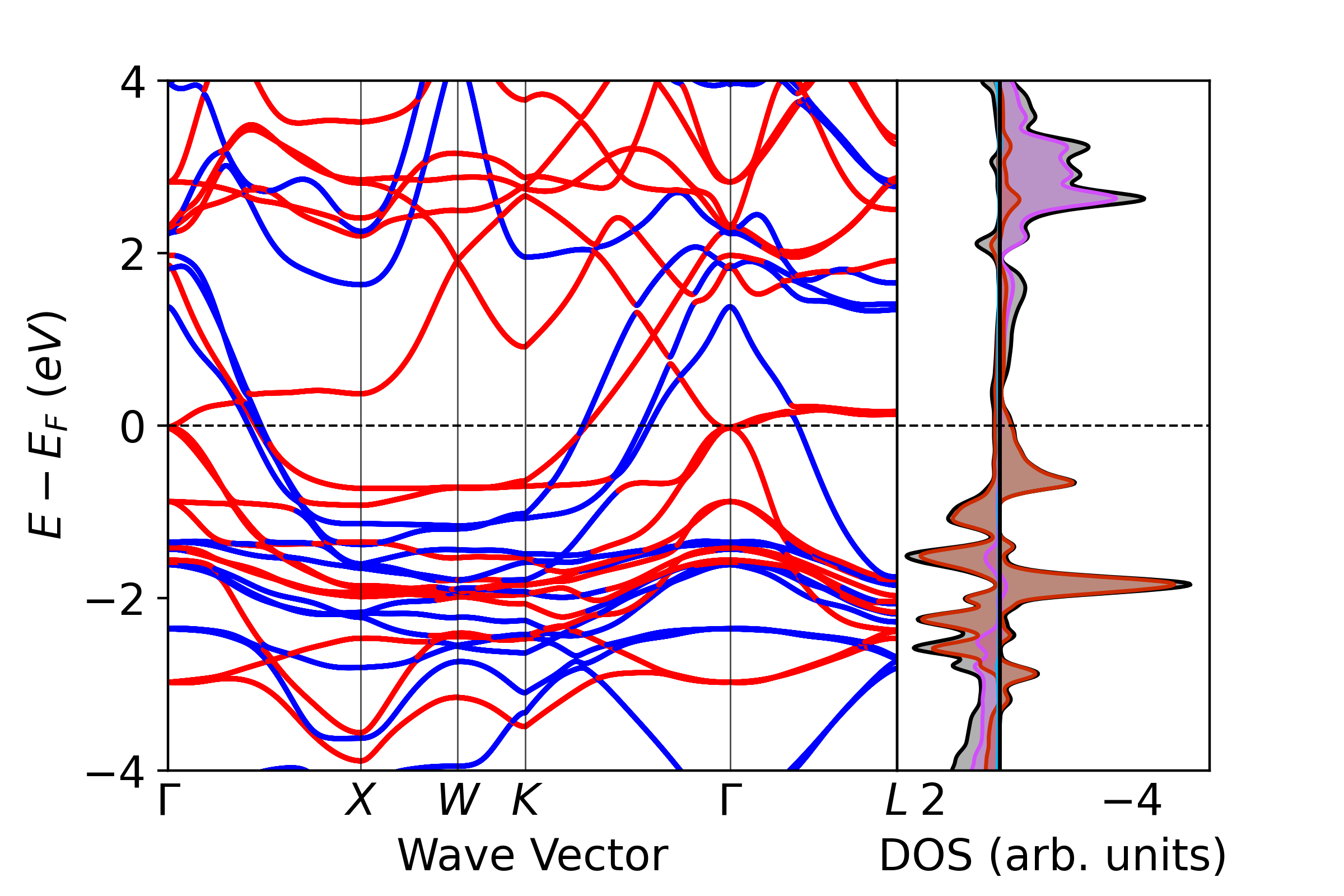}\\ (b) QP$GW$ (reference)
\end{minipage}\\[3pt]
\begin{minipage}[t]{0.31\textwidth}\centering
\includegraphics[width=\linewidth]{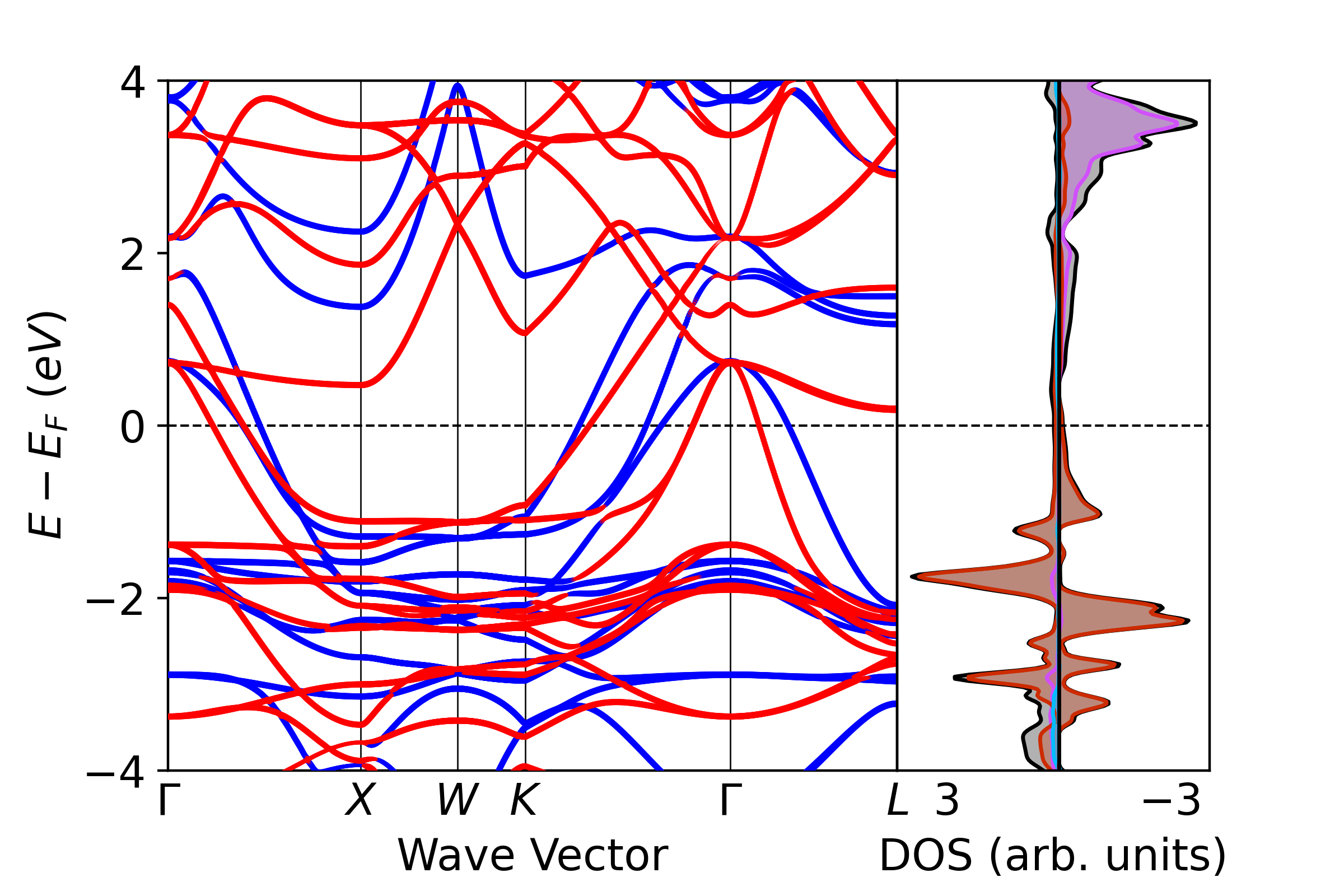}\\ (c) PBE+$U$(LR)
\end{minipage}\hspace{1em}
\begin{minipage}[t]{0.31\textwidth}\centering
\includegraphics[width=\linewidth]{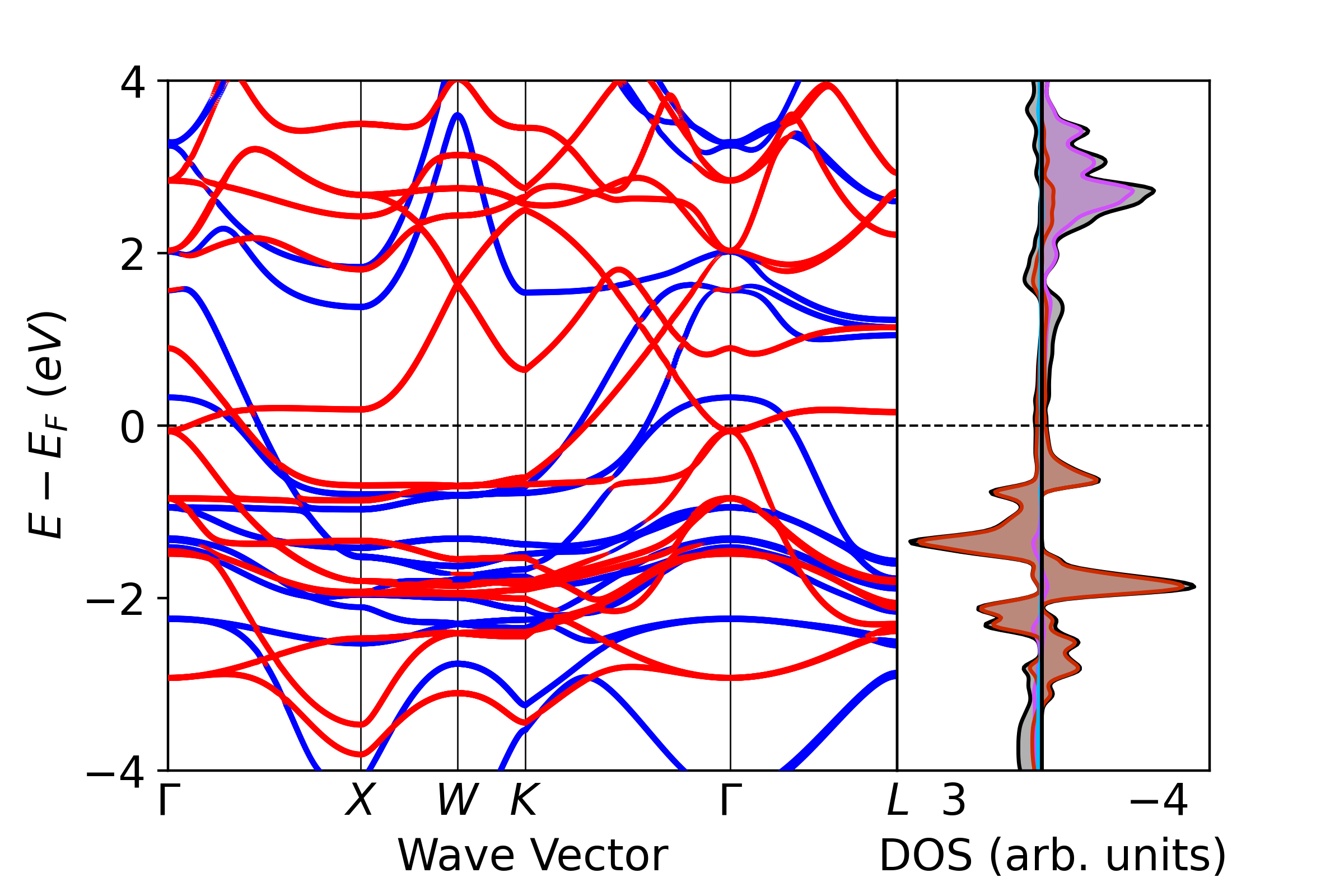}\\ (d) PBE+$U$(BO)
\end{minipage}
\caption{Band structure and element-resolved DOS of Ni$_2$MnSb calculated using (a) PBE, (b) QP$GW$ (the reference), (c) PBE+$U$(LR) ($U_{\mathrm{Ni}}=5.33$~eV, $U_{\mathrm{Mn}}=6.67$~eV), and (d) PBE+$U$(BO) ($U_{\mathrm{Ni}}=-1.51$~eV, $U_{\mathrm{Mn}}=5.16$~eV). In the band-structure panels, the majority and minority spin channels are shown in blue and red, respectively. The large positive LR $U$ values over-localize the transition-metal $d$ states and enlarge their exchange splitting, distorting the electronic structure away from the QP$GW$ reference, whereas the negative BO $U_{\mathrm{Ni}}$ keeps the itinerant Ni $3d$ states delocalized, as in QP$GW$.}
\label{fig:LRbands_NiMnSb}
\end{figure}

\begin{figure}[H]
\def\baselinestretch{1}\footnotesize
\centering
\begin{minipage}[t]{0.31\textwidth}\centering
\includegraphics[width=\linewidth]{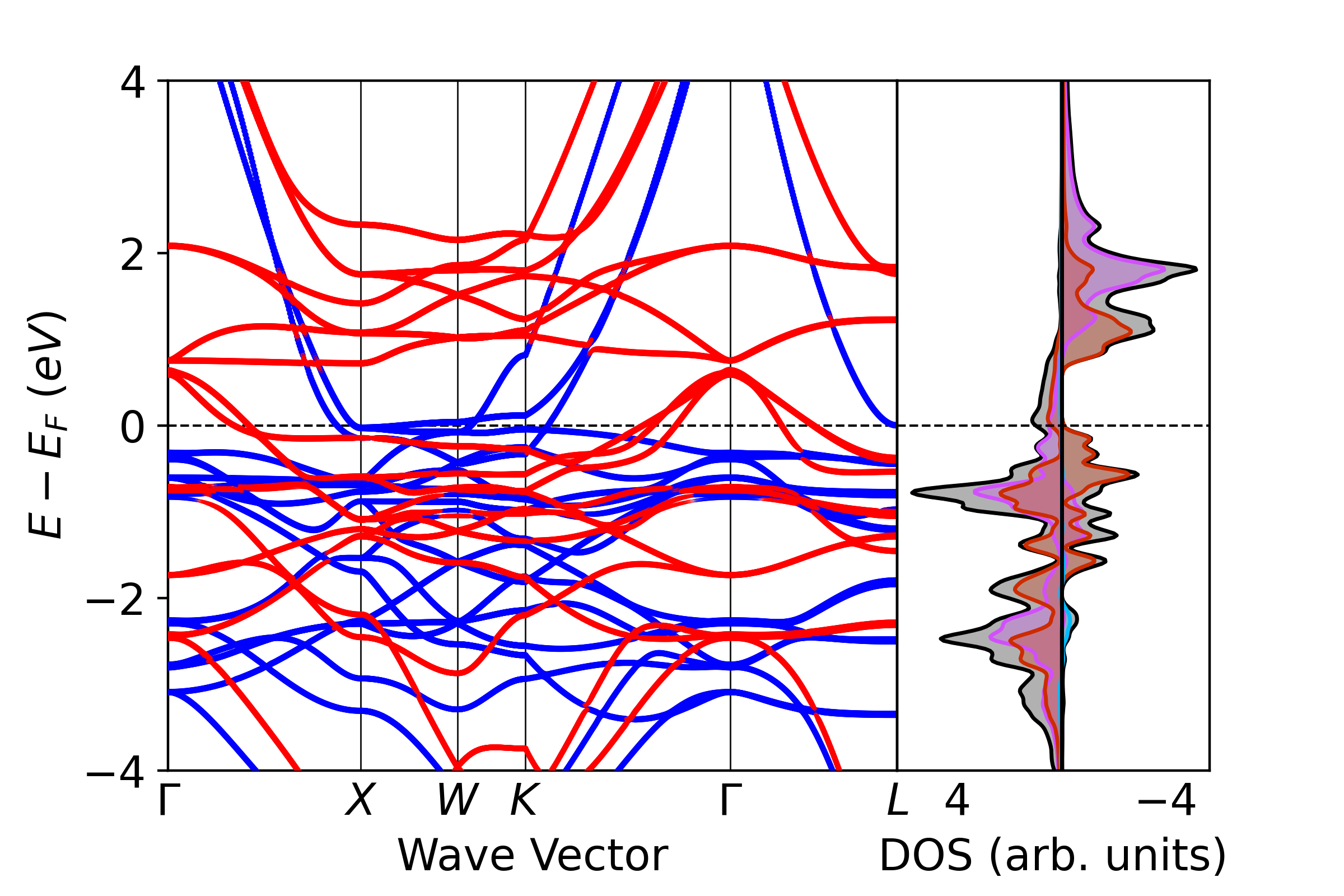}\\ (a) PBE
\end{minipage}\hspace{1em}
\begin{minipage}[t]{0.31\textwidth}\centering
\includegraphics[width=\linewidth]{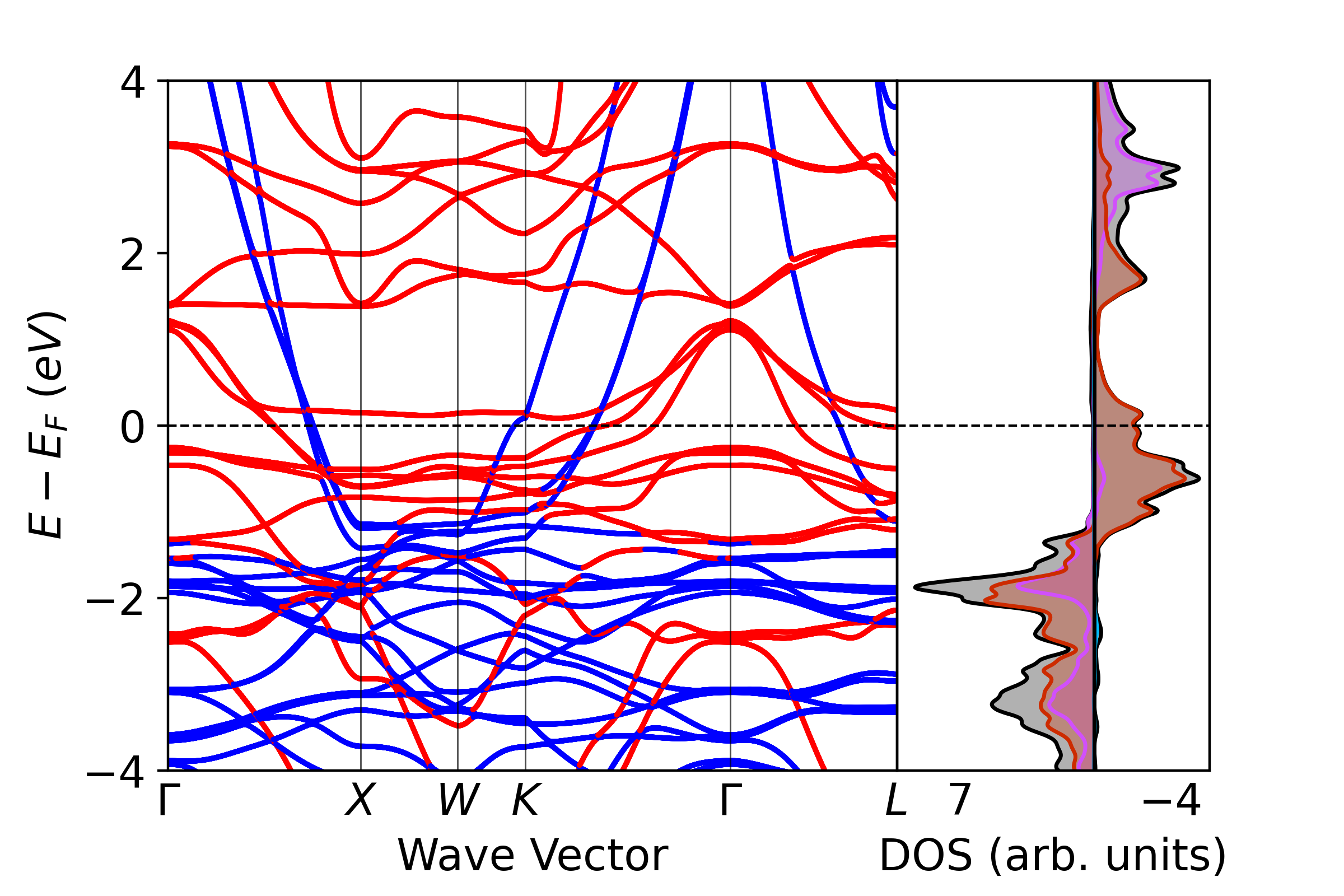}\\ (b) QP$GW$ (reference)
\end{minipage}\\[3pt]
\begin{minipage}[t]{0.31\textwidth}\centering
\includegraphics[width=\linewidth]{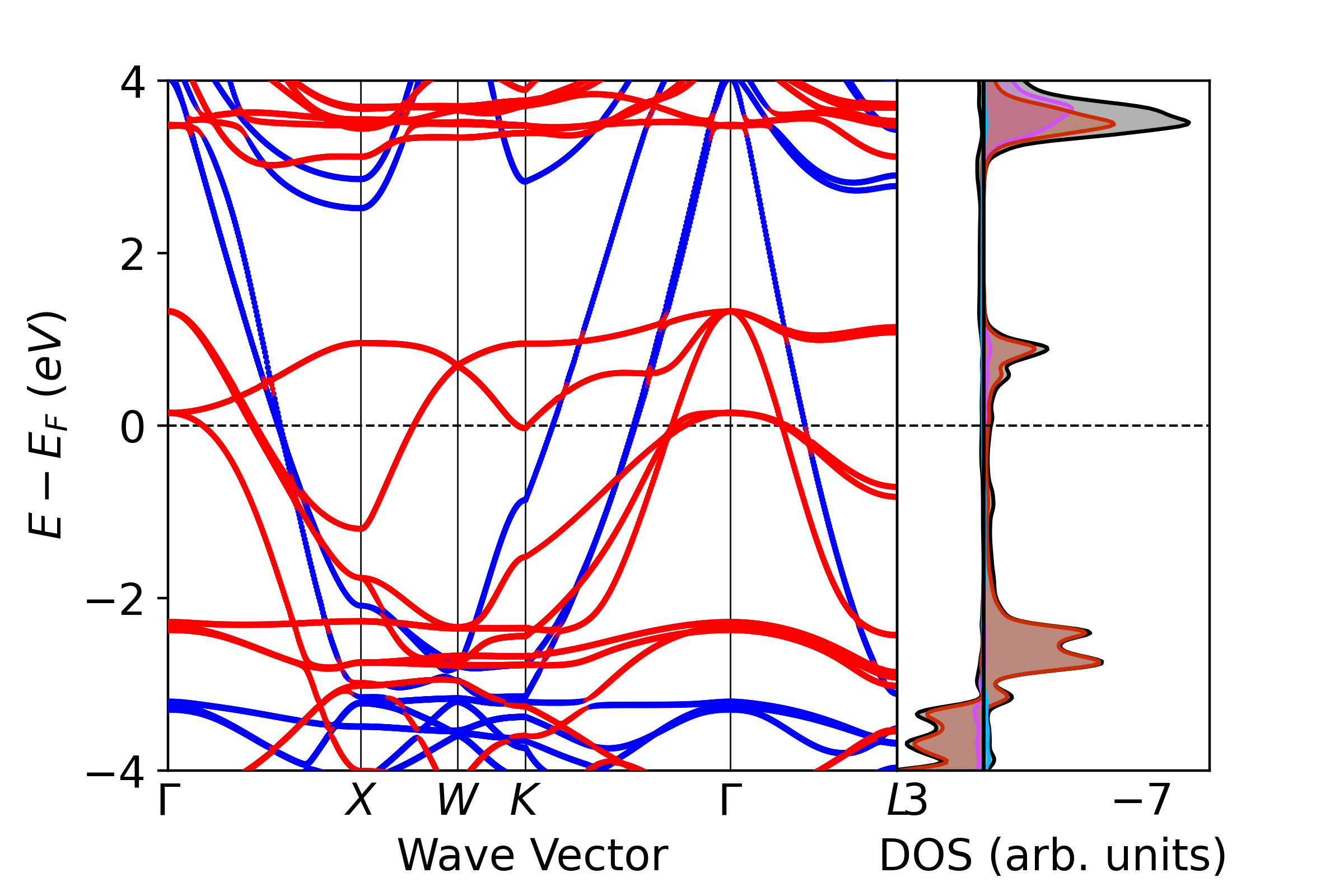}\\ (c) PBE+$U$(LR)
\end{minipage}\hspace{1em}
\begin{minipage}[t]{0.31\textwidth}\centering
\includegraphics[width=\linewidth]{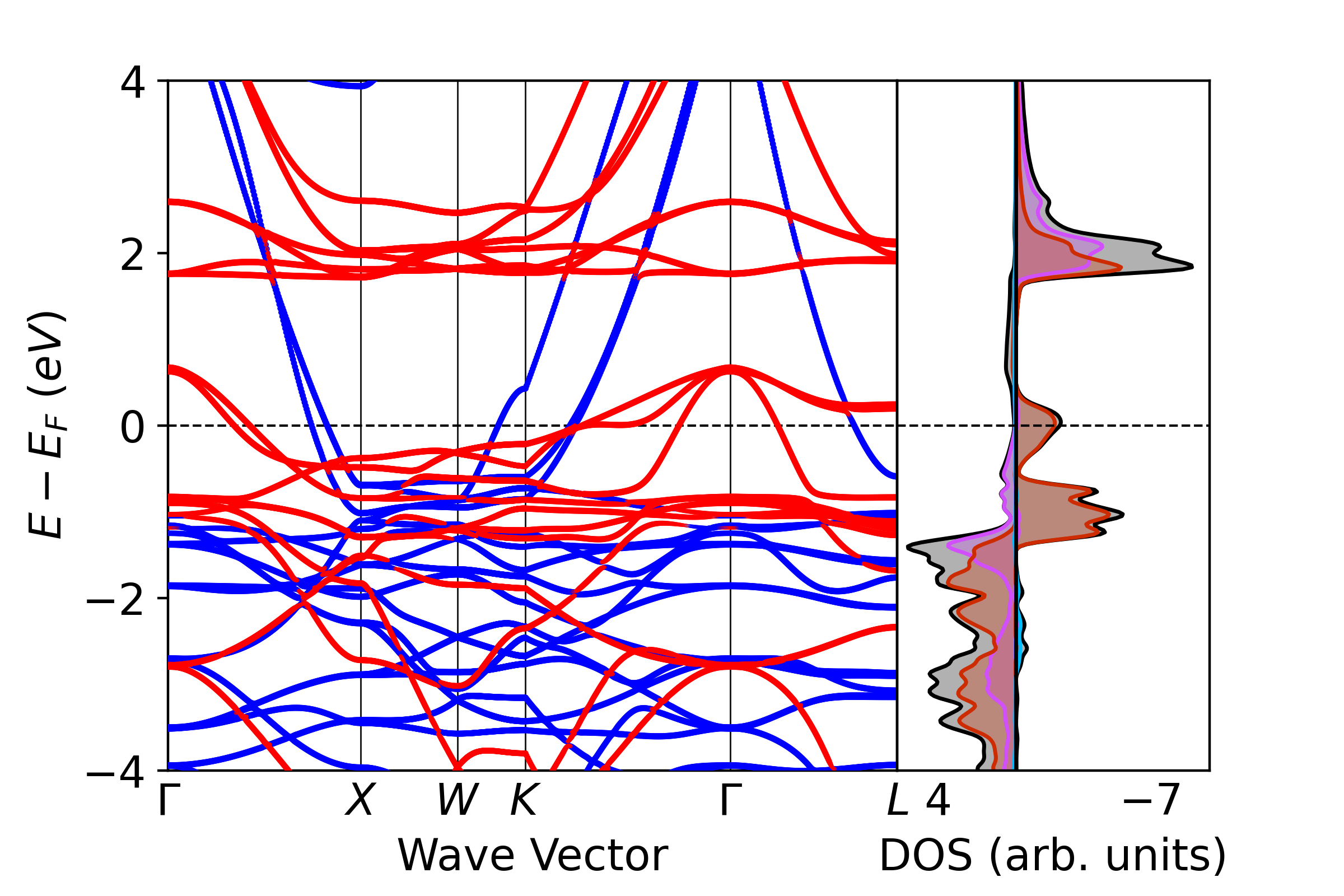}\\ (d) PBE+$U$(BO)
\end{minipage}
\caption{Same layout as Fig.~\ref{fig:LRbands_NiMnSb} for Co$_2$MnIn, with (c) PBE+$U$(LR) ($U_{\mathrm{Co}}=5.27$~eV, $U_{\mathrm{Mn}}=6.83$~eV) and (d) PBE+$U$(BO) ($U_{\mathrm{Co}}=2.25$~eV, $U_{\mathrm{Mn}}=0.75$~eV). The large positive LR $U$ over-localizes the Mn $d$ manifold, whereas the small BO $U$ keeps the Mn $d$ states positioned as in QP$GW$.}
\label{fig:LRbands_Co2MnIn}
\end{figure}

Table~\ref{tab:LR} compares the LR and BO $U$ values, their accuracy with respect to the QP$GW$ reference, and their computational cost. In contrast to the strongly system- and site-dependent $U$(BO), the LR approach yields nearly uniform, large positive values across the six compounds ($\sim$5~eV on the X site and $\sim$6--7~eV on Mn). These large corrections over-localize the $d$ states and enlarge their exchange splitting, pushing the electronic structure toward the HSE limit, which overestimates the exchange splitting for these materials, and away from QP$GW$, which lies between PBE and HSE (see the main text). This is consistent with the previously reported tendency of the LR approach to overestimate $U$ for the delocalized Co $d$ states in Co-based Heuslers \cite{2019_Nawa_DFTUCoBasedHeusler_RSCAdv}. As a result, PBE+$U$(BO) reproduces the QP$GW$ band structures and magnetic moments more closely than PBE+$U$(LR) for every compound, with respect to both descriptors ($\Delta$Band and $\Delta$Mag, defined in the main text), and significantly so on average (mean $\Delta$Band of 0.53 vs.\ 0.90~eV and mean $\Delta$Mag of 0.14 vs.\ 0.43~$\mu_B$). We also note that the LR approach cannot, by construction, yield a negative $U_{\mathrm{eff}}$, whereas the QP$GW$-optimal correction for Ni in Ni$_2$MnSb is negative (see the main text). For Ni$_2$MnSb, the negative $U^{\mathrm{BO}}_{\mathrm{Ni}} = -1.51$~eV keeps the itinerant Ni $3d$ states delocalized, as in QP$GW$ ($\Delta$Band $= 0.40$~eV, $\Delta$Mag $= 0.09~\mu_B$), whereas the large positive $U^{\mathrm{LR}}_{\mathrm{Ni}} = 5.33$~eV over-localizes the Ni $d$ manifold and enlarges its exchange splitting, driving the electronic structure away from QP$GW$ ($\Delta$Band $= 0.70$~eV, $\Delta$Mag $= 0.18~\mu_B$) [Fig.~\ref{fig:LRbands_NiMnSb}]. Figures~\ref{fig:LRbands_NiMnSb} and \ref{fig:LRbands_Co2MnIn} compare the PBE, QP$GW$, PBE+$U$(LR), and PBE+$U$(BO) band structures and DOS of two representative compounds, Ni$_2$MnSb and Co$_2$MnIn.

In terms of computational cost, $C^{\mathrm{BO}}$ comprises the BO optimization loop plus the one-time QP$GW$ reference, and $C^{\mathrm{LR}}$ is the full eight-point LR determination in the $3\times3\times3$ supercell. All timings were obtained on NERSC Perlmutter and are reported in CPU-node-hour equivalents. On average, the converged LR determination is $\sim$2.5$\times$ more expensive than the entire BO workflow, including its QP$GW$ reference. Moreover, the QP$GW$ reference is a one-time cost per material: the resulting $U$(BO) may subsequently be transferred to surfaces, interfaces, and large supercells at no additional reference cost.

\begin{table*}[h]
\centering
\caption{Bayesian-optimization versus linear-response Hubbard $U$ (in eV, on the X/Y transition-metal sites), accuracy relative to the QP$GW$ reference, and computational cost. $\Delta_b$ is the band-structure RMSE (in eV) over the 16 bands nearest $E_F$ and $\Delta_m$ is the per-atom magnetic-moment RMSE (in $\mu_B$), both measured against QP$GW$ (smaller = closer to QP$GW$). Costs $C$ are in CPU-node-hour equivalents; $C^{\mathrm{BO}}$ includes the one-time QP$GW$ reference.}
\label{tab:LR}
\vspace{6pt}
\setlength{\tabcolsep}{10pt}
\begin{tabular}{lcccccccc}
\toprule
 & $U^{\mathrm{BO}}$ (X/Y) & $U^{\mathrm{LR}}$ (X/Y)
 & $\Delta_b^{\mathrm{BO}}$ & $\Delta_b^{\mathrm{LR}}$
 & $\Delta_m^{\mathrm{BO}}$ & $\Delta_m^{\mathrm{LR}}$
 & $C^{\mathrm{BO}}$ & $C^{\mathrm{LR}}$ \\
\midrule
Co$_2$MnIn & $2.25/0.75$  & $5.27/6.83$ & 0.63 & 1.47 & 0.18 & 0.44 & 49.7 & 166.9 \\
Co$_2$MnSn & $1.51/1.65$  & $5.05/6.69$ & 0.46 & 1.19 & 0.18 & 0.51 & 54.5 & 163.5 \\
Co$_2$TiSn & $3.53/3.65$  & $4.63/6.21$ & 0.57 & 0.58 & 0.26 & 0.45 & 65.4 & 104.2 \\
Co$_2$ZrAl & $1.33/9.38$  & $5.07/3.46$ & 0.62 & 0.66 & 0.05 & 0.74 & 63.2 & 112.6 \\
Ni$_2$MnSb & $-1.51/5.16$ & $5.33/6.67$ & 0.40 & 0.70 & 0.09 & 0.18 & 51.7 & 126.9 \\
Ni$_2$MnSn & $2.39/2.35$  & $5.39/6.87$ & 0.51 & 0.81 & 0.10 & 0.24 & 51.0 & 177.5 \\
\midrule
mean       &              &             & 0.53 & 0.90 & 0.14 & 0.43 & 55.9 & 141.9 \\
\bottomrule
\end{tabular}
\end{table*}

\clearpage

\clearpage
\section{Orbital-Resolved Band Structures}

\begin{figure*}[h]
\centering
\includegraphics[width=1\textwidth]{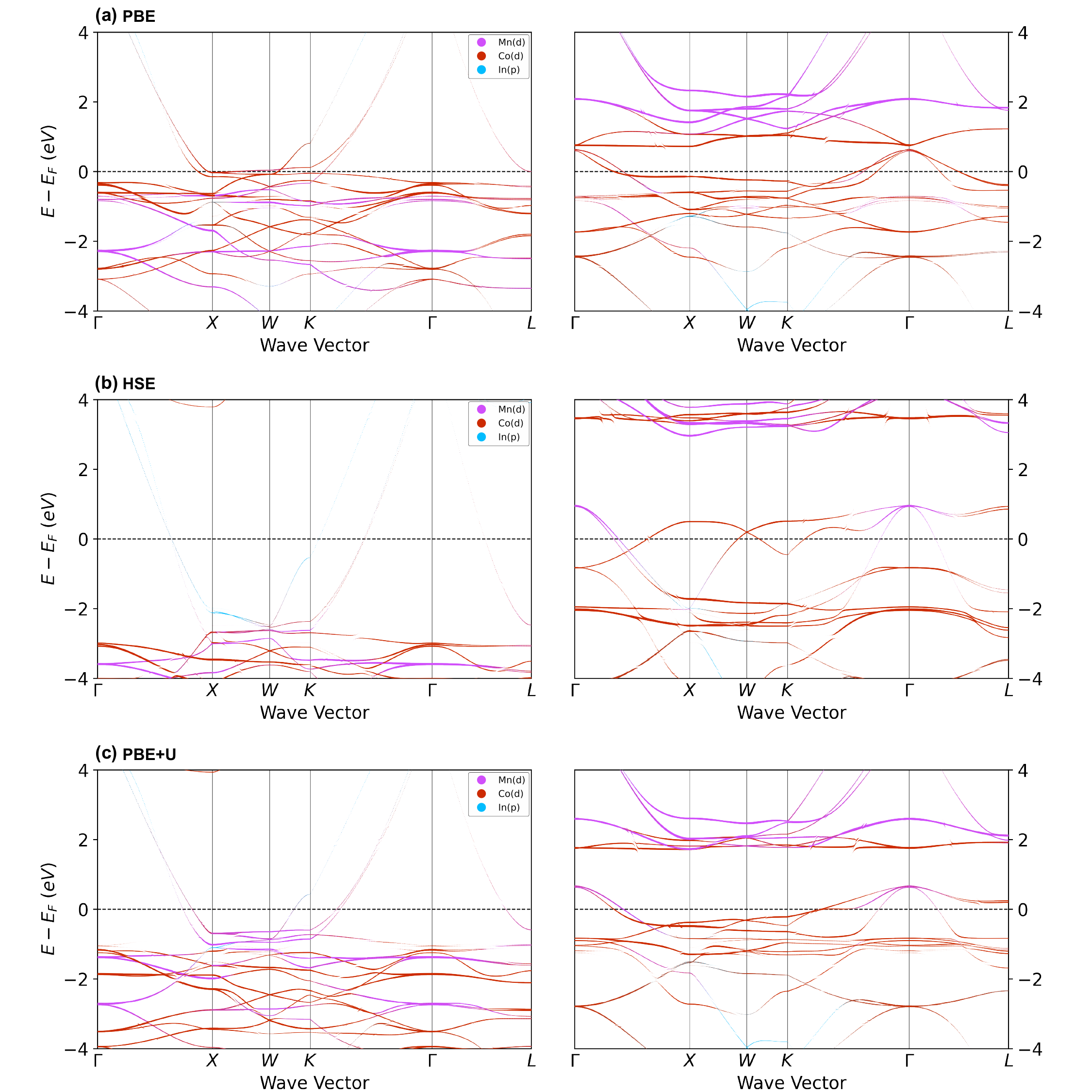}
\caption{Orbital-resolved band structures of Co$_2$MnIn for the majority-spin (left) and minority-spin (right) channels, calculated using (a) PBE, (b) HSE, and (c) PBE+$U$(BO).}
\end{figure*}

\begin{figure*}[h]
\centering
\includegraphics[width=1\textwidth]{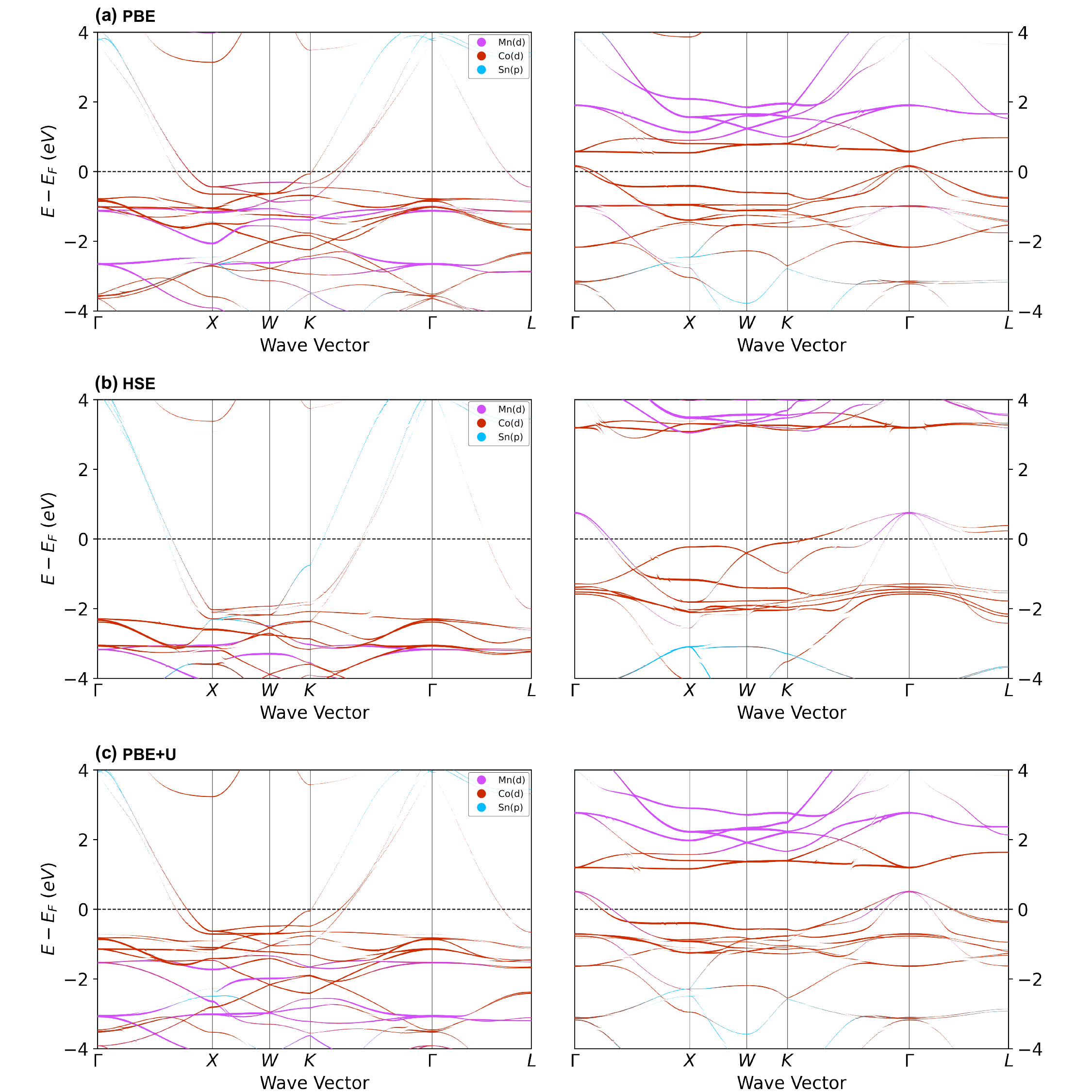}
\caption{Orbital-resolved band structures of Co$_2$MnSn for the majority-spin (left) and minority-spin (right) channels, calculated using (a) PBE, (b) HSE, and (c) PBE+$U$(BO).}
\end{figure*}

\begin{figure*}[h]
\centering
\includegraphics[width=1\textwidth]{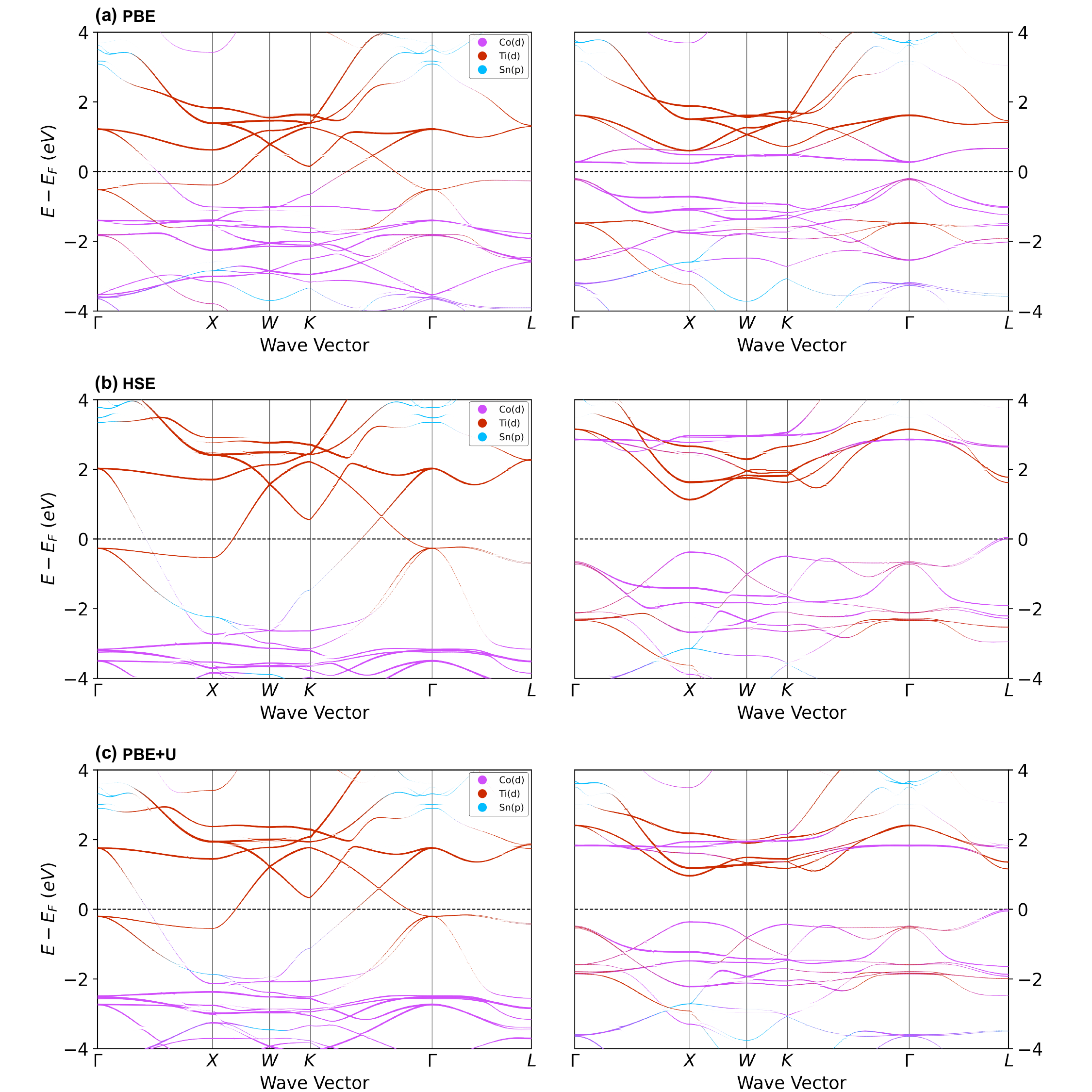}
\caption{Orbital-resolved band structures of Co$_2$TiSn for the majority-spin (left) and minority-spin (right) channels, calculated using (a) PBE, (b) HSE, and (c) PBE+$U$(BO).}
\end{figure*}

\begin{figure*}[h]
\centering
\includegraphics[width=1\textwidth]{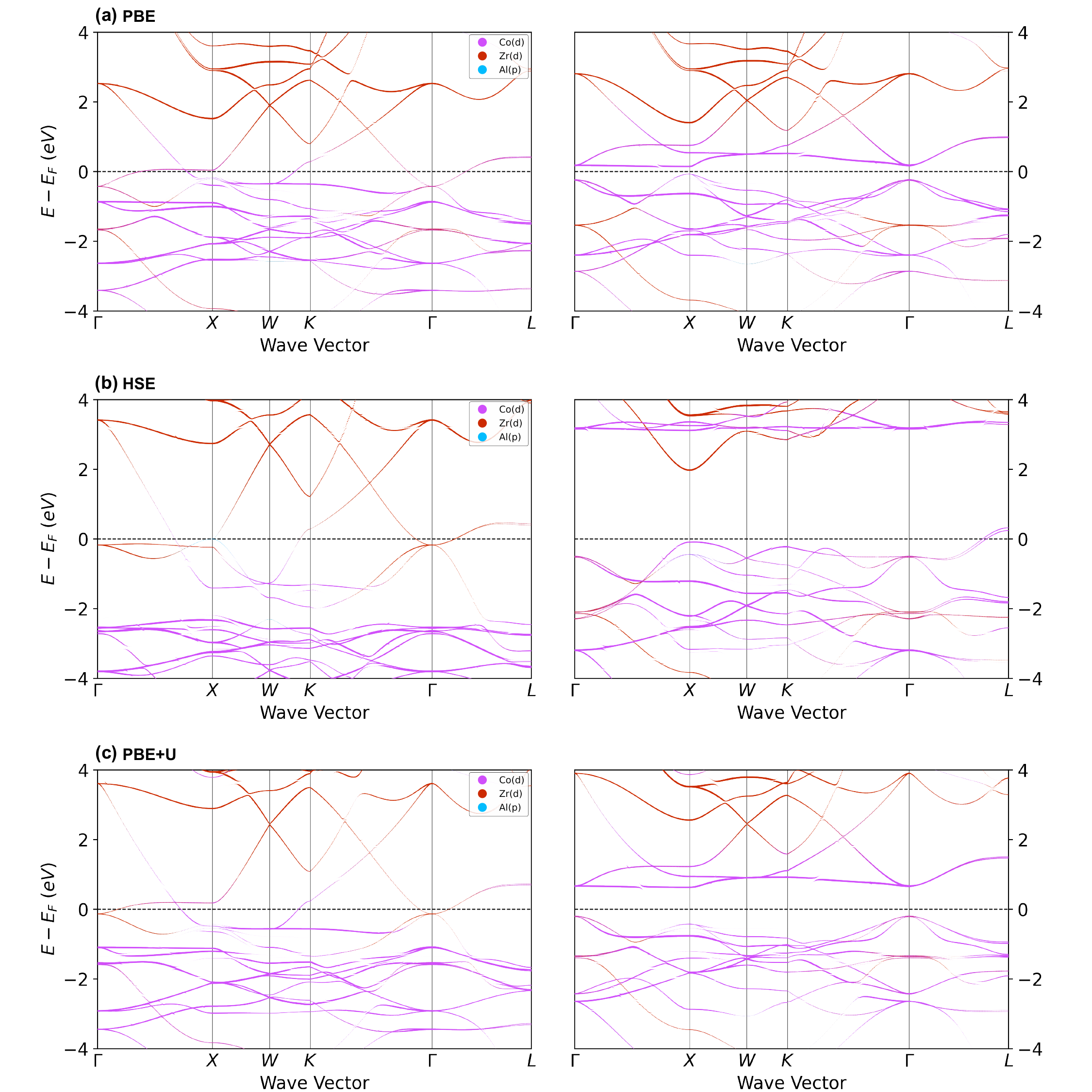}
\caption{Orbital-resolved band structures of Co$_2$ZrAl for the majority-spin (left) and minority-spin (right) channels, calculated using (a) PBE, (b) HSE, and (c) PBE+$U$(BO).}
\label{fig:orb_Co2ZrAl}
\end{figure*}

\begin{figure*}[h]
\centering
\includegraphics[width=1\textwidth]{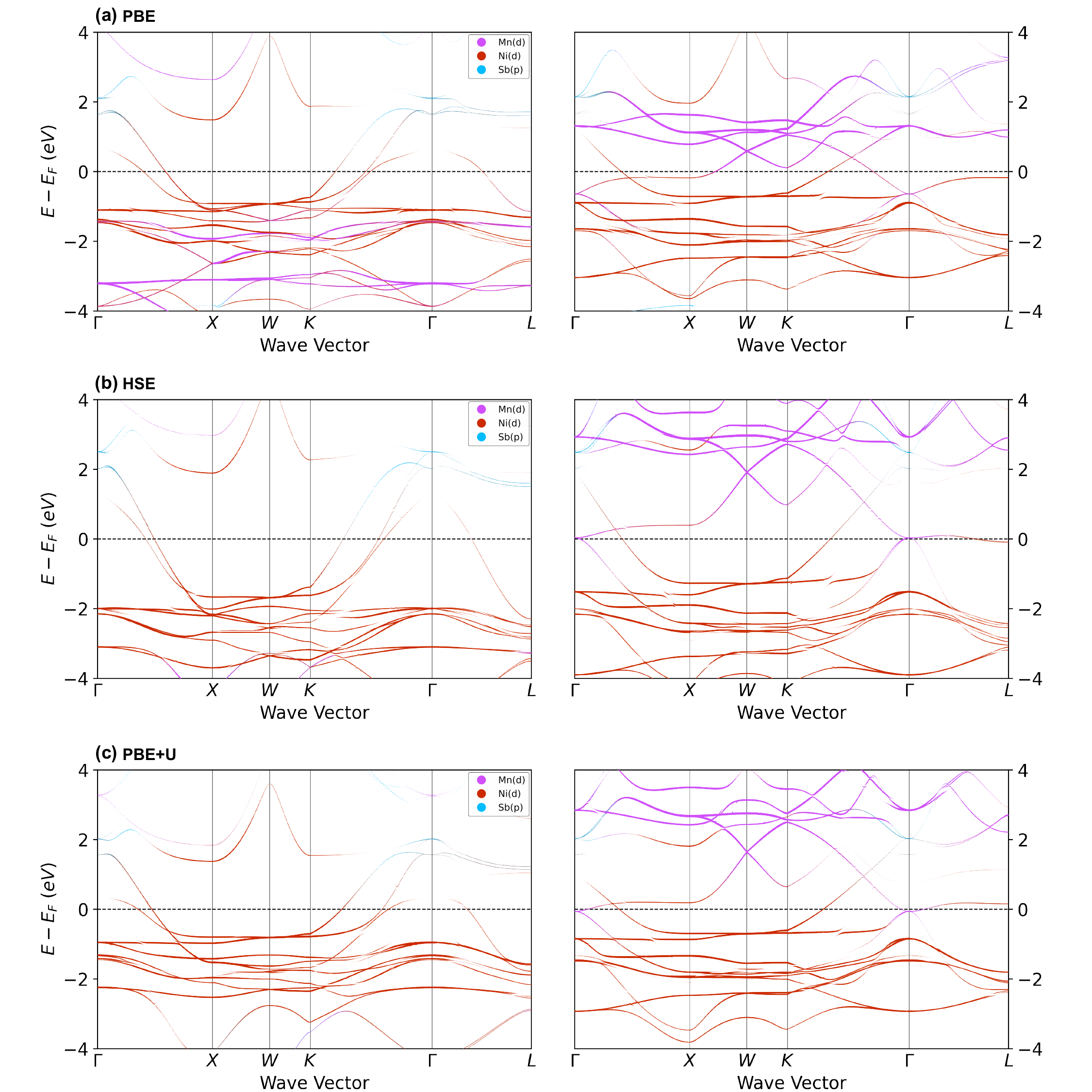}
\caption{Orbital-resolved band structures of Ni$_2$MnSb for the majority-spin (left) and minority-spin (right) channels, calculated using (a) PBE, (b) HSE, and (c) PBE+$U$(BO).}
\end{figure*}

\begin{figure*}[h]
\centering
\includegraphics[width=1\textwidth]{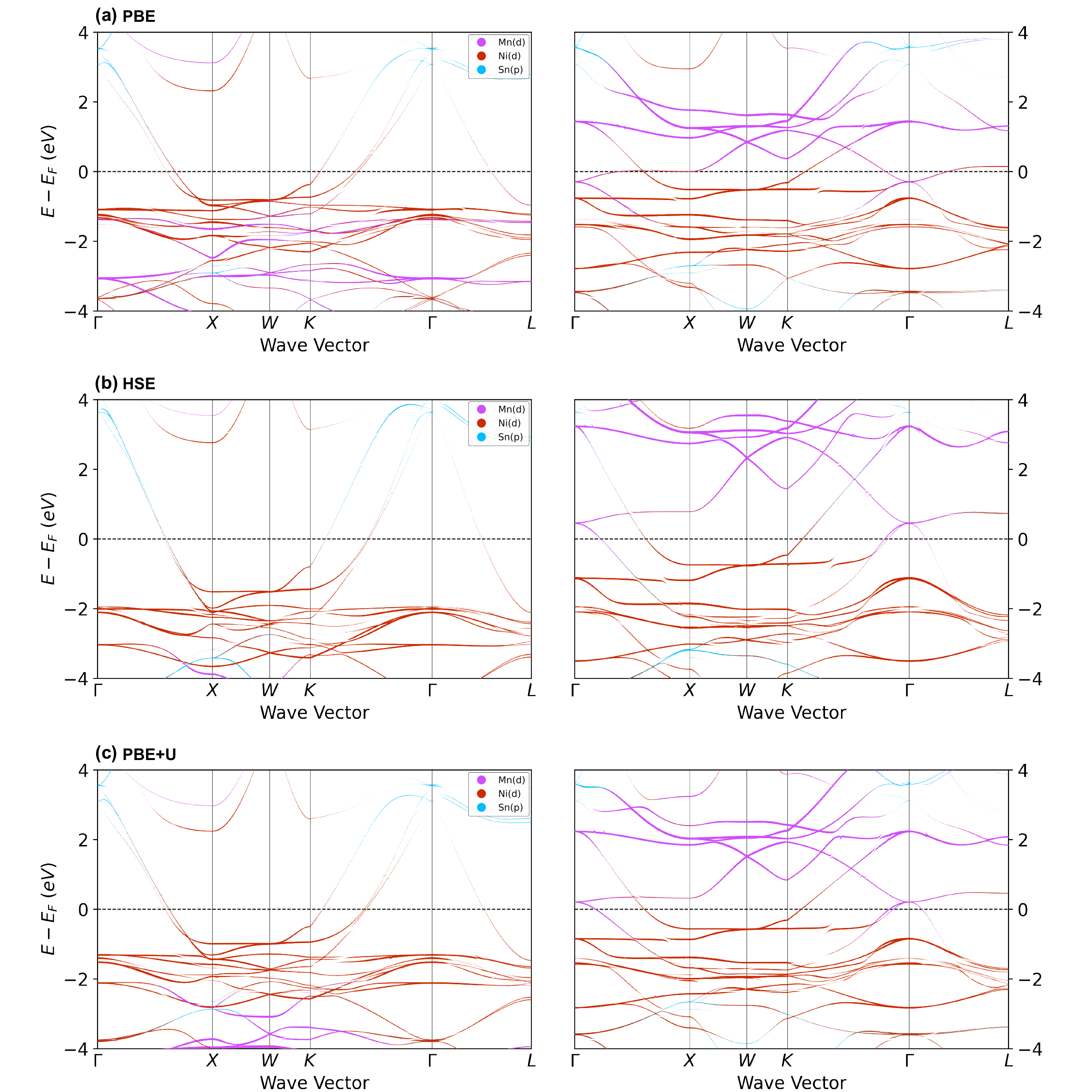}
\caption{Orbital-resolved band structures of Ni$_2$MnSn for the majority-spin (left) and minority-spin (right) channels, calculated using (a) PBE, (b) HSE, and (c) PBE+$U$(BO).}
\end{figure*}

\clearpage
\section{BO Convergence}

\begin{figure*}[h]
\centering
\includegraphics[width=1\textwidth]{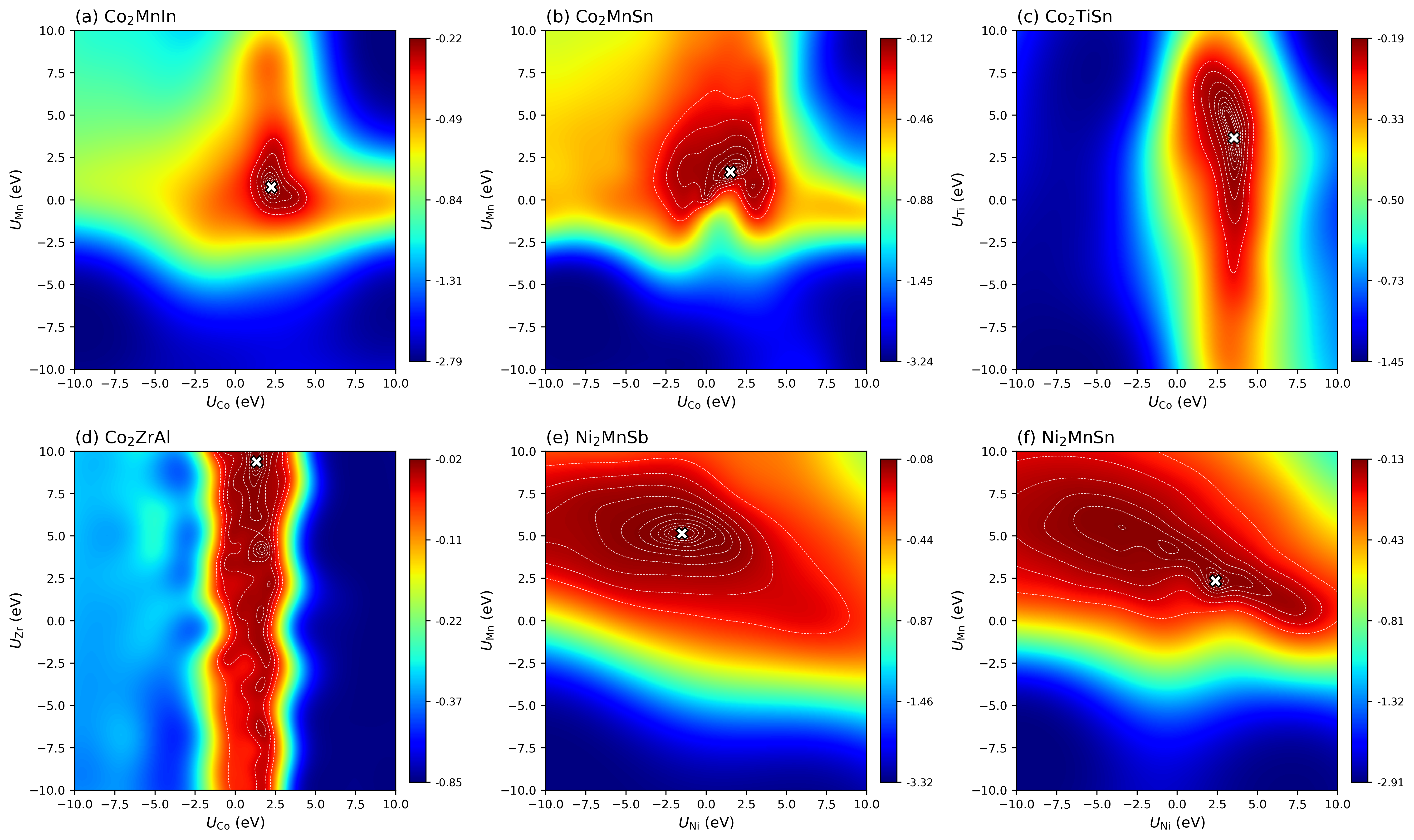}
\caption{Landscape of the Bayesian optimization objective function (Gaussian-process-predicted mean of the objective defined in Eq.\ 1, with $\alpha_1 = 0.0$, $\alpha_2 = 0.5$, and $\alpha_3 = 0.5$) across the Hubbard $U$ range of [-10, 10] eV for:
(a) Co$_2$MnIn ($U^{\text{opt}}_\text{Co}=2.252$\,eV, $U^{\text{opt}}_\text{Mn}=0.751$\,eV),
(b) Co$_2$MnSn ($U^{\text{opt}}_\text{Co}=1.512$\,eV, $U^{\text{opt}}_\text{Mn}=1.652$\,eV),
(c) Co$_2$TiSn ($U^{\text{opt}}_\text{Co}=3.534$\,eV, $U^{\text{opt}}_\text{Ti}=3.654$\,eV),
(d) Co$_2$ZrAl ($U^{\text{opt}}_\text{Co}=1.331$\,eV, $U^{\text{opt}}_\text{Zr}=9.379$\,eV),
(e) Ni$_2$MnSb ($U^{\text{opt}}_\text{Ni}=-1.512$\,eV, $U^{\text{opt}}_\text{Mn}=5.155$\,eV),
and (f) Ni$_2$MnSn ($U^{\text{opt}}_\text{Ni}=2.392$\,eV, $U^{\text{opt}}_\text{Mn}=2.352$\,eV). The color map uses a nonlinear scale that enhances the contrast near the maximum, and white iso-contours are drawn in the top objective band. The white crosses indicate the optimal Hubbard $U$ values, corresponding to the maximum of the objective function; each landscape exhibits a well-defined maximum, except for Co$_2$ZrAl, whose landscape forms a ridge that is sharp along $U_{\text{Co}}$ but flat along $U_{\text{Zr}}$. The flat direction corresponds to a $d$ shell that is nearly empty and far from the Fermi level, to which the objective function is physically insensitive.
}
\label{fig:obj}
\end{figure*}

\begin{figure*}[h]
\centering
\includegraphics[width=1\textwidth]{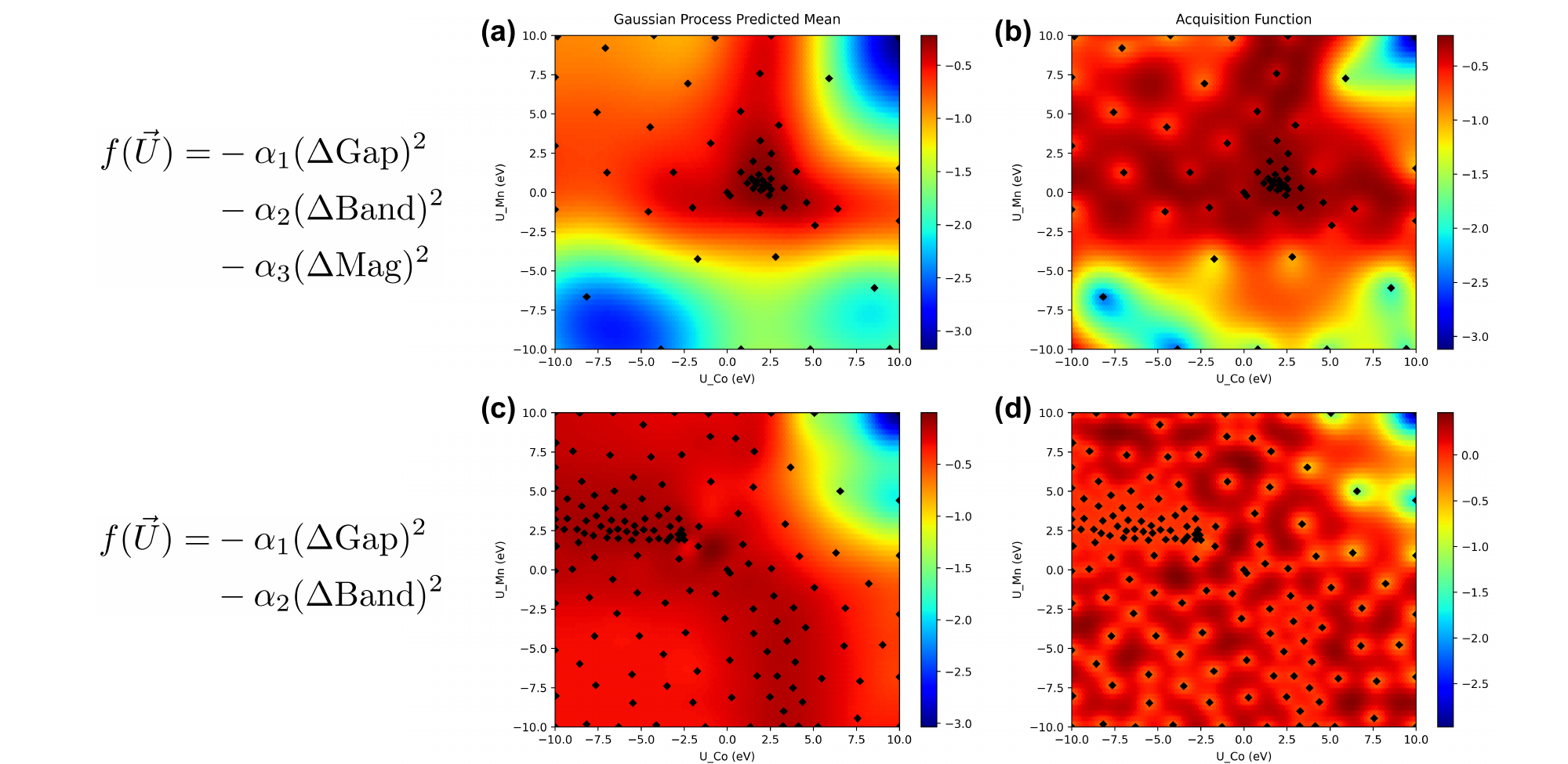}
\caption{Gaussian process-predicted mean and acquisition function for the Bayesian optimization (BO) of the Hubbard $U$ parameter for Co$_2$MnIn. (a)-(b) show results using the updated objective function, which includes contributions from atomic magnetic moments. (c)-(d) show results using the previous objective function \cite{yu2020machine} based solely on band gap and eigenvalue differences. In both cases, $\alpha_1$ = 0 (metallic system) and $\alpha_2$ = 0.5, with $\alpha_3$ = 0.5 only in the updated case. The magnetic moment-enhanced BO exhibits a distinct local maximum within the range of [-10, 10] eV, whereas the band-only BO produces a relatively flat objective function landscape. This result demonstrates the effectiveness of including the magnetic term, leading to regression toward a more informative Hubbard $U$, increased convergence speed, and robustness against data variations and noise.
}
\label{fig:compare}
\end{figure*}

\clearpage
\section{1D Versus 2D Optimization of the Hubbard $U$: The Case of \NoCaseChange{Co$_2$ZrAl}}\label{sec:UZr}

The Bayesian optimization landscape of Co$_2$ZrAl in Fig.~\ref{fig:obj}(d) forms a ridge that is sharp along $U_{\mathrm{Co}}$ but nearly flat along $U_{\mathrm{Zr}}$. The physical origin of this flatness can be deduced from the orbital-resolved band structures in Fig.~\ref{fig:orb_Co2ZrAl}: Zr in Co$_2$ZrAl is nominally $4d^0$, and the Zr $4d$ spectral weight in PBE [Fig.~\ref{fig:orb_Co2ZrAl}(a)] is concentrated in the conduction bands, roughly 1~eV and higher above the Fermi level, whereas the occupied manifold is dominated by Co $3d$ states. Occupation-dependent quantities, in particular the atomic magnetic moments, are therefore expected to be insensitive to $U_{\mathrm{Zr}}$. It should be noted, however, that the positions of the unoccupied Zr-$4d$-derived bands---and through them the band-structure term of the objective function---do respond to it.

\begin{figure*}[h]
\centering
\includegraphics[width=1\textwidth]{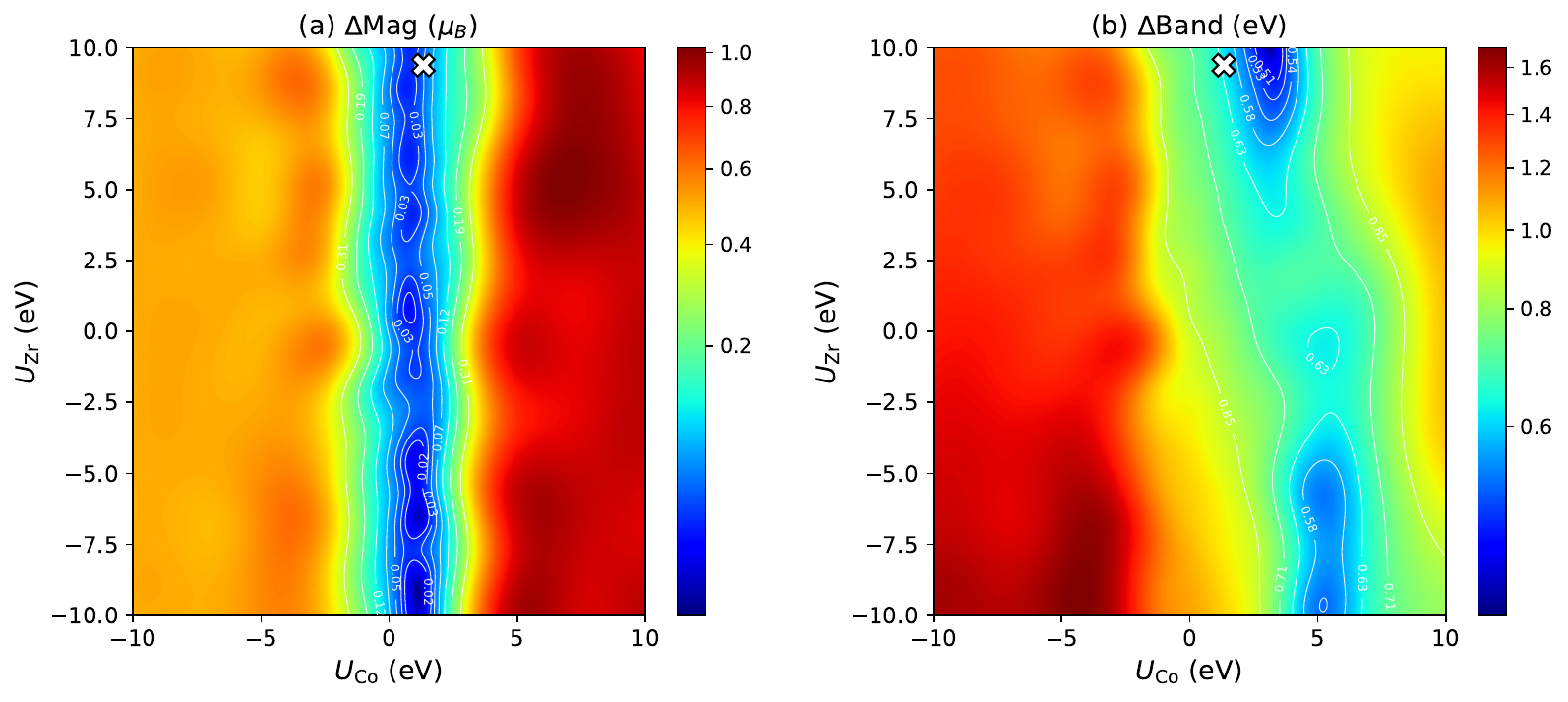}
\caption{Gaussian-process-predicted mean of the two error metrics entering the objective function for Co$_2$ZrAl, fitted separately to the data set of the two-dimensional optimization: (a) the magnetic-moment difference $\Delta$Mag and (b) the band-structure difference $\Delta$Band with respect to the QP$GW$ reference. The color maps use a nonlinear scale that enhances the contrast near the minimum, white iso-contours are drawn in the low-error band, and the white cross marks the optimal ($U_{\mathrm{Co}}$, $U_{\mathrm{Zr}}$). The $\Delta$Mag iso-contours run nearly parallel to the $U_{\mathrm{Zr}}$ axis, showing that the magnetic moments are governed by $U_{\mathrm{Co}}$ alone, whereas $\Delta$Band is lowered by a positive $U_{\mathrm{Zr}}$, which tunes the unoccupied Zr $4d$ bands.}
\label{fig:deltamaps}
\end{figure*}

\begin{figure*}[h]
\centering
\begin{minipage}[t]{0.42\textwidth}\centering
\includegraphics[width=\linewidth]{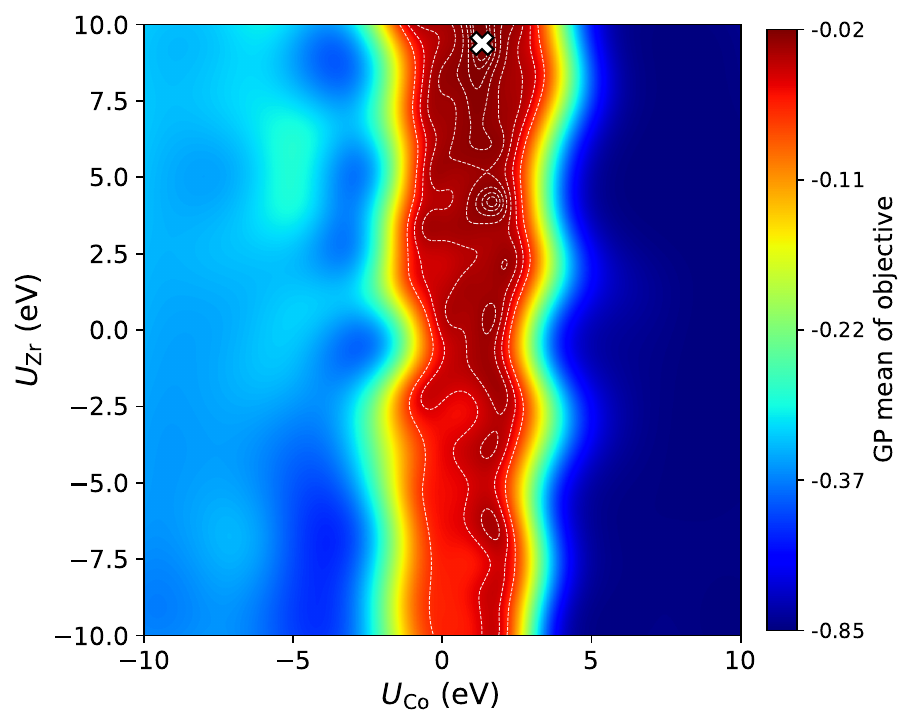}\\ (a) 2D: $U_{\mathrm{Co}}$ and $U_{\mathrm{Zr}}$
\end{minipage}\hspace{1em}
\begin{minipage}[t]{0.42\textwidth}\centering
\includegraphics[width=\linewidth]{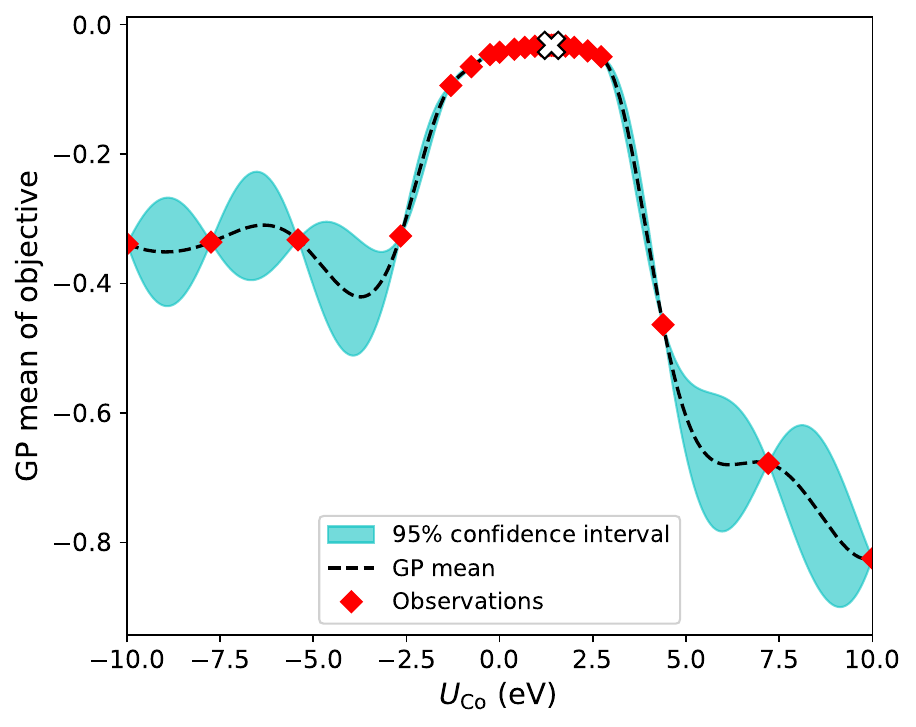}\\ (b) 1D: $U_{\mathrm{Co}}$ only
\end{minipage}\\[6pt]
\begin{minipage}[t]{0.42\textwidth}\centering
\includegraphics[width=\linewidth]{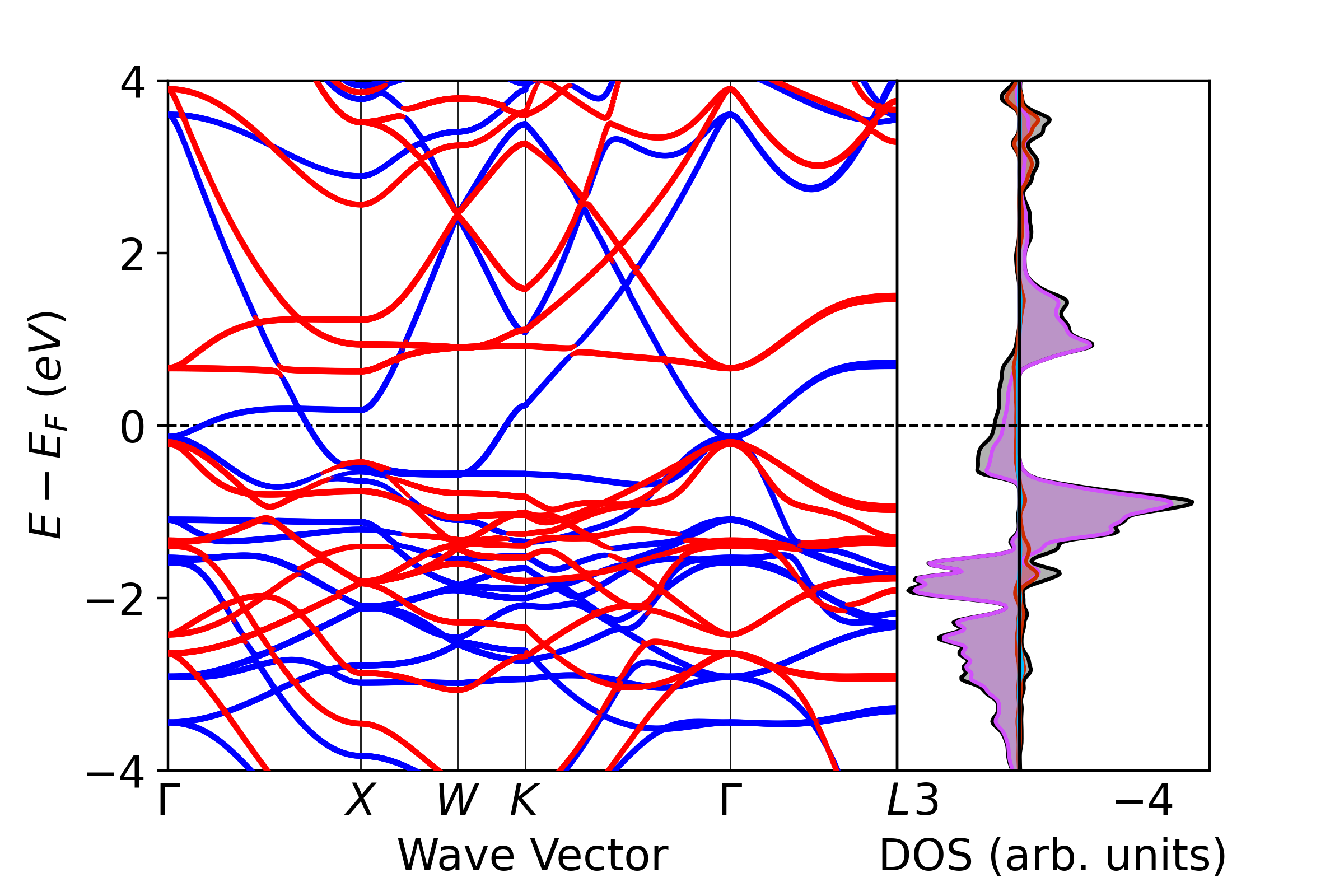}\\ (c) PBE+$U$(BO), 2D optimum
\end{minipage}\hspace{1em}
\begin{minipage}[t]{0.42\textwidth}\centering
\includegraphics[width=\linewidth]{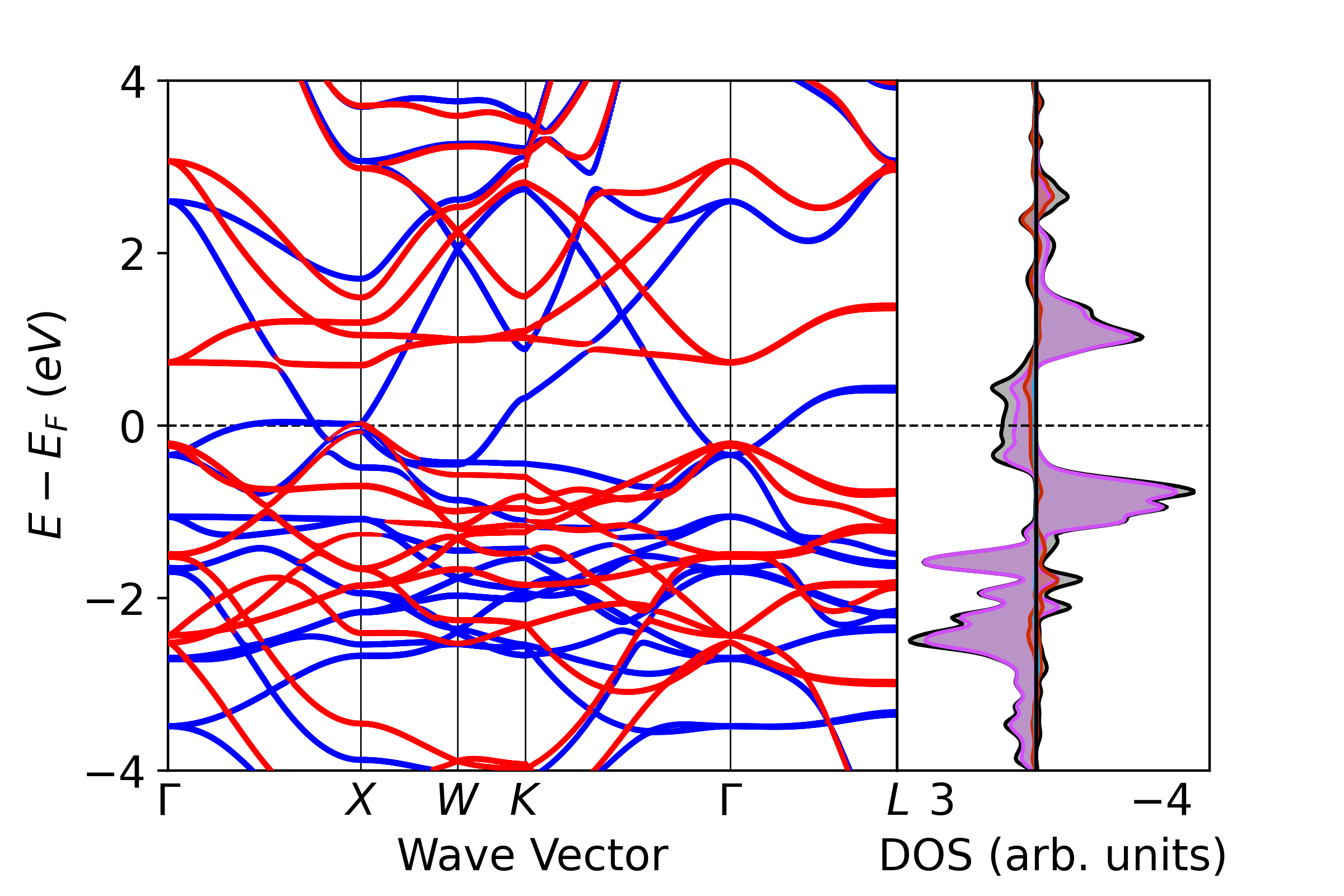}\\ (d) PBE+$U$(BO), 1D optimum
\end{minipage}
\caption{Bayesian optimization of the Hubbard $U$ for Co$_2$ZrAl in two and one dimensions. (a) Gaussian-process-predicted mean of the objective function over the ($U_{\mathrm{Co}}$, $U_{\mathrm{Zr}}$) plane, in the same representation as Fig.~\ref{fig:obj}; the white cross marks the optimum ($U_{\mathrm{Co}} = 1.33$~eV, $U_{\mathrm{Zr}} = 9.38$~eV). (b) Gaussian-process-predicted mean (dashed line) and 95\% confidence interval for the one-dimensional optimization of $U_{\mathrm{Co}}$ with $U_{\mathrm{Zr}} = 0$~eV; red diamonds mark the sampled points and the cross the optimum ($U_{\mathrm{Co}} = 1.39$~eV). (c),(d) PBE+$U$(BO) band structures and element-resolved DOS obtained with the two- and one-dimensional optimal $U$ values, respectively; the majority and minority spin channels are shown in blue and red. The two solutions are nearly indistinguishable around the Fermi level and differ mainly in the position of the unoccupied Zr-$4d$-derived conduction bands (cf.\ Fig.~\ref{fig:orb_Co2ZrAl}).}
\label{fig:1d2d}
\end{figure*}

To confirm this, we repeated the optimization for Co$_2$ZrAl in one dimension, applying $U$ to Co only (i.e., fixing $U_{\mathrm{Zr}} = 0$~eV), with all other settings identical. Fig.~\ref{fig:deltamaps} shows Gaussian-process maps of the two error metrics, $\Delta$Band and $\Delta$Mag, extracted from the 2D optimization, and Fig.~\ref{fig:1d2d} shows the Gaussian-process-predicted mean of the objective function for the two optimizations, together with the PBE+$U$(BO) band structures and DOS obtained at the respective optima. Three observations emerge. First, $\Delta$Mag is essentially unaffected by $U_{\mathrm{Zr}}$: its iso-contours in Fig.~\ref{fig:deltamaps}(a) run nearly parallel to the $U_{\mathrm{Zr}}$ axis, and the optimized values are 0.047~$\mu_B$ (2D) and 0.037~$\mu_B$ (1D). Second, the band-structure agreement benefits from including $U_{\mathrm{Zr}}$: $\Delta$Band improves from 0.79~eV without it to 0.62~eV with it, as the unoccupied Zr-$4d$ bands are shifted toward their QP$GW$ positions [Fig.~\ref{fig:deltamaps}(b)]. Third, the optimal Co value is robust: $U_{\mathrm{Co}} = 1.33$~eV in 2D versus $1.39$~eV in 1D. Other aspects of the electronic structure near the Fermi level, including the half-metallic character and the total magnetization, are unchanged between the two approaches [Figs.~\ref{fig:1d2d}(c) and \ref{fig:1d2d}(d)].

Including $U_{\mathrm{Zr}}$ in the optimization is therefore the correct choice when agreement with the QP$GW$ reference is sought for both the band structure and the magnetic moments. Only if the unoccupied bands were not of interest would the 1D $U_{\mathrm{Co}}$-only optimization be sufficient (it converges within 30 iterations, a quarter of the 2D cost). For Co$_2$ZrAl, however, the unoccupied bands are important: the magnitude of the minority-spin gap that defines half-metallicity depends on their position, and the device applications of half-metals, such as magnetic tunnel junctions and spin injection into semiconductors \cite{peterson2016spin, palmstrom2016heusler, datta1990electronic, holub2007electrical}, involve transport through unoccupied states in an energy window above the Fermi level set by the applied voltage. Hence, the two-dimensional optimization is the appropriate choice for all the materials studied here.

\clearpage
\section{QP$GW$ Convergence}
\begin{figure*}[h]
\centering
\includegraphics[width=1\textwidth]{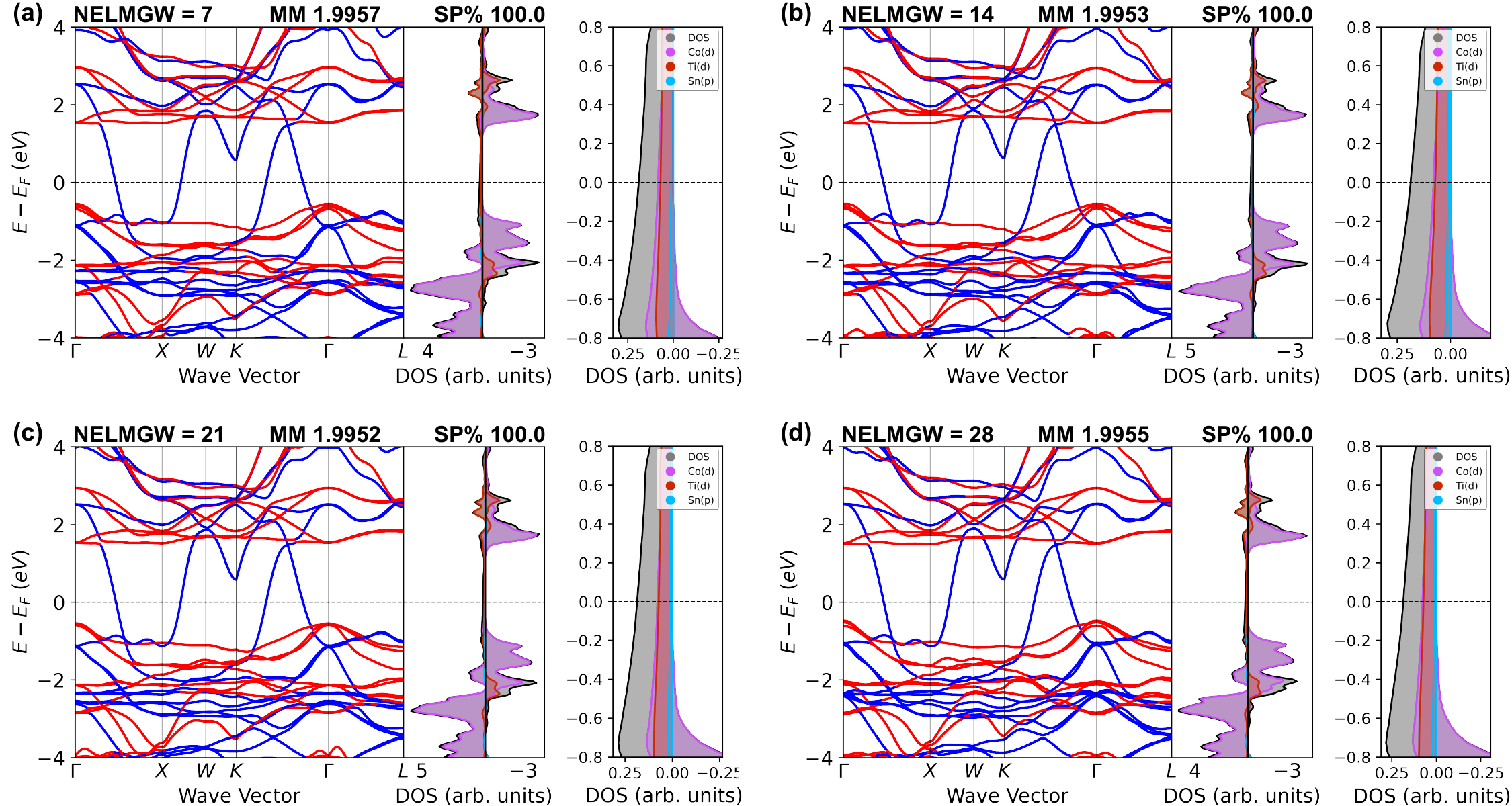}
\caption{Convergence tests for self-consistent QP$GW$ calculations illustrated by the band structures and density of states (DOS) of Co$_2$TiSn obtained with a varying number of QP$GW$ iterations, i.e., the number of cycles in which quasiparticle energies and orbitals are updated self-consistently in both $G$ and $W$. Results are shown for (a) \texttt{NELMGW} = 7, (b) \texttt{NELMGW} = 14, (c) \texttt{NELMGW} = 21, and (d) \texttt{NELMGW} = 28. From 7 to 28 iterations, only small changes occur in the deeper valence states below $E_F - 2~\mathrm{eV}$ in the band structure, accompanied by negligible variations in the total magnetic moment. Moreover, the spin polarization at the Fermi level remains 100\%. Notably, Co$_2$TiSn contains the largest number of valence electrons (within the PAW pseudopotential framework) among the six materials studied here and is therefore expected to exhibit the slowest convergence behavior. These results justify our choice of using seven self-consistent QP$GW$ iterations for all calculations reported in this work.
}
\end{figure*}

\clearpage
\section{Effect of SOC}

\begin{figure*}[h]
\centering
\includegraphics[width=1\textwidth]{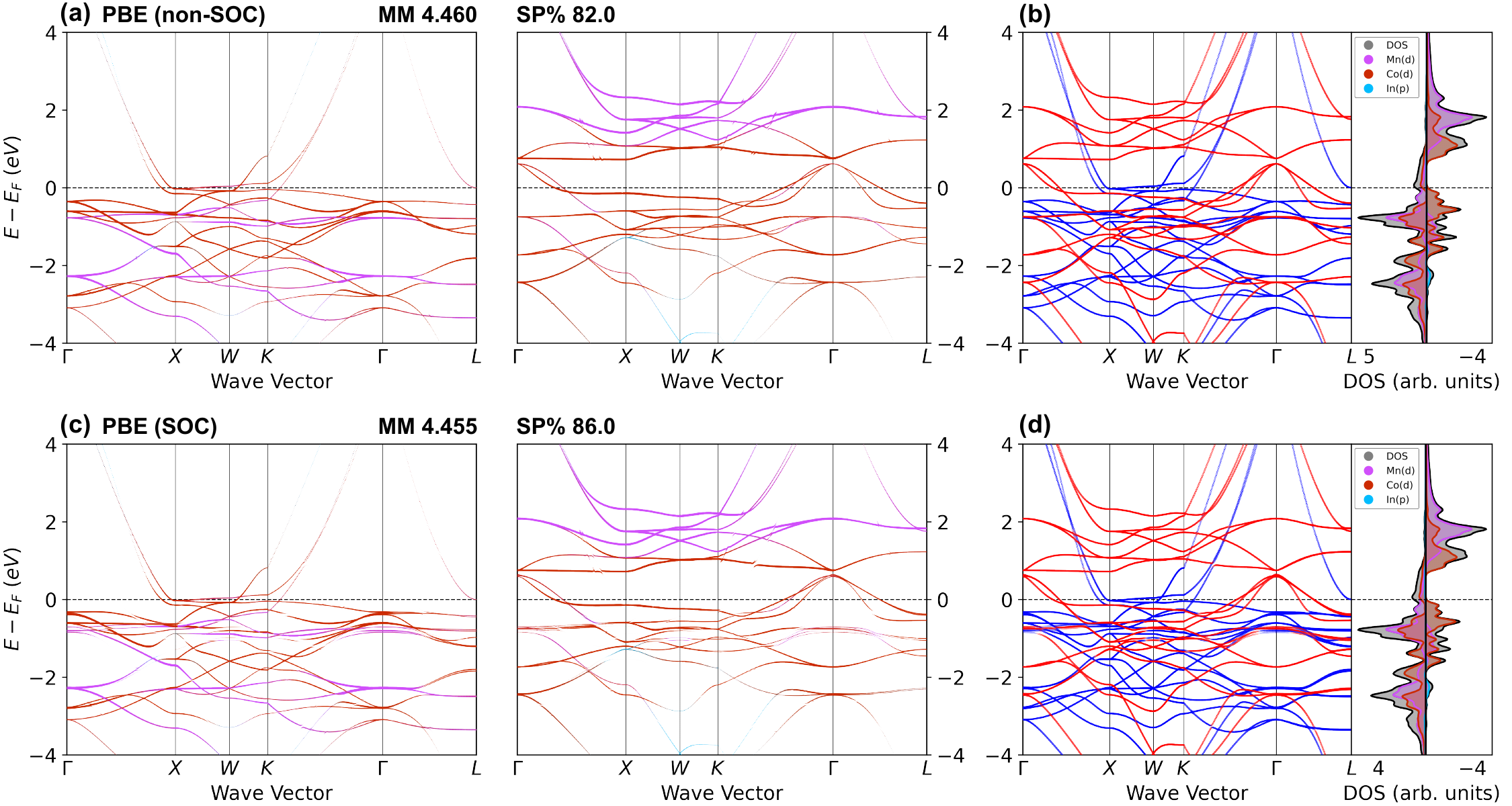}
\caption{Band structures and DOS of Co$_2$MnIn calculated using the PBE functional: (a,b) without spin-orbit coupling (SOC) and (c,d) with SOC. Panels (a) and (c) display the orbital-resolved band structures for the majority-spin (left) and minority-spin (right) channels, while panels (b) and (d) present the spin-resolved band structures and orbital-resolved DOS. The majority- and minority-spin channels are shown in blue and red, respectively. Inclusion of SOC induces pronounced spin-orbit-driven band inversions, particularly near the $\Gamma$ and W points. It also leads to avoided crossings, most evident along the $\Gamma$-W path. These effects result in changes in the spin polarization and slight variations in the total magnetic moment. Similar behavior is observed across all six systems; consequently, SOC is included in all calculations for the Heusler compounds studied here.}
\end{figure*}

\begin{figure*}[h]
\centering
\includegraphics[width=1\textwidth]{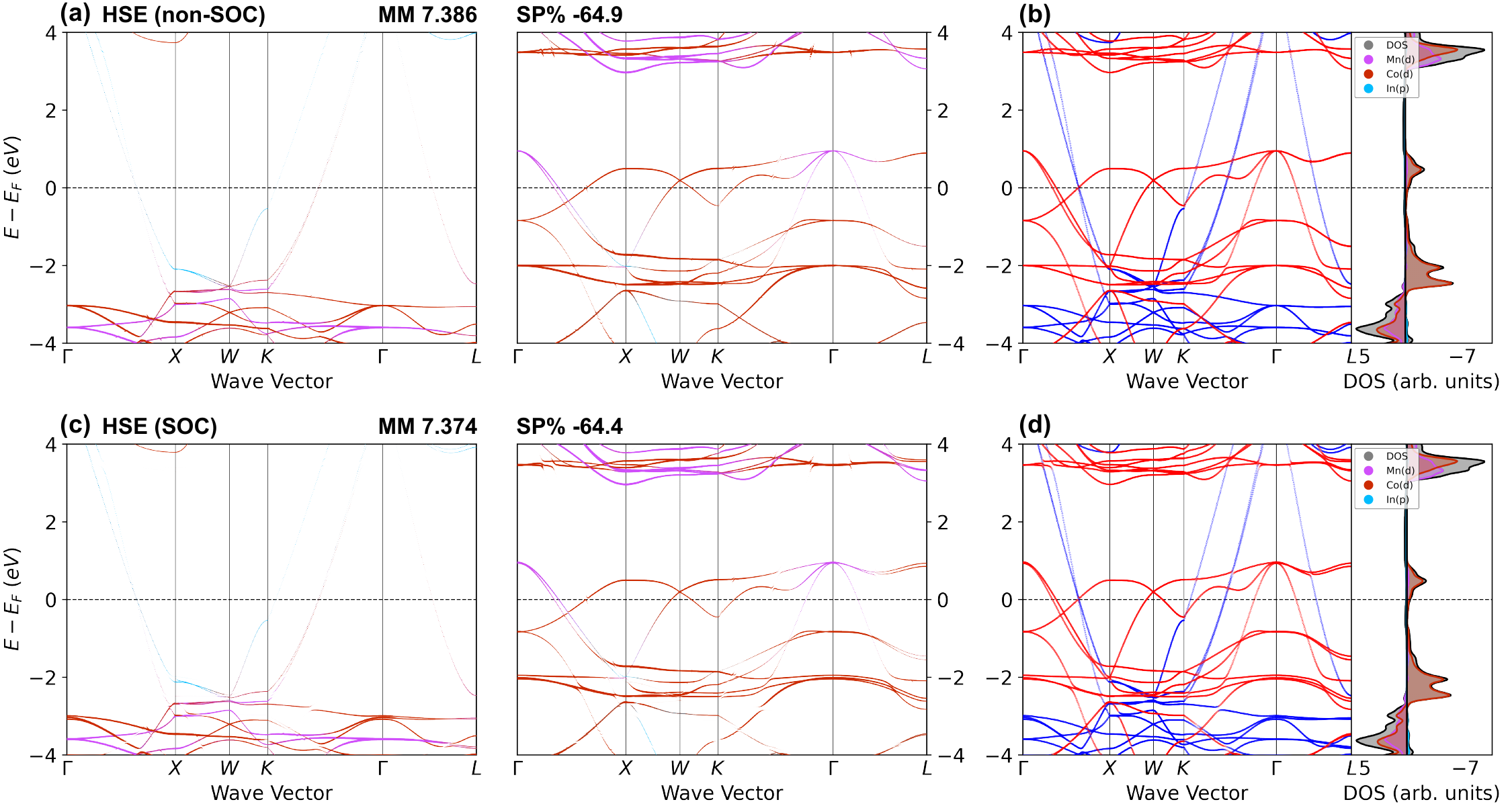}
\caption{Band structures and DOS of Co$_2$MnIn calculated using the HSE functional: (a,b) without spin-orbit coupling (SOC) and (c,d) with SOC. Panels (a) and (c) display the orbital-resolved band structures for the majority-spin (left) and minority-spin (right) channels. Panels (b) and (d) present the spin-resolved band structures and orbital-resolved DOS. The majority- and minority-spin channels are shown in blue and red, respectively. Inclusion of SOC induces pronounced spin-orbit-driven band inversions, particularly near the $\Gamma$ and W points. It also leads to avoided crossings, most evident along the $\Gamma$-W path. These effects result in slight changes in the spin polarization and the total magnetic moment.}
\end{figure*}

\clearpage
\section{Spin Polarization Sensitivity Analysis}

\begin{table*}[h]
\centering
\caption{Sensitivity of the spin polarization (SP, in \%) of Co$_2$MnIn to the size of the integration window $2\Delta$ (in eV) around the Fermi level. The SP values reported in the main text are obtained by integrating over an energy window of $2\Delta = 3.5\,k_B T = 0.091\,\mathrm{eV}$ centered at the Fermi level, corresponding to the FWHM of the derivative of the Fermi–Dirac distribution at room temperature. The results are robust to the choice of integration window.}
\label{tab:SP_window_sensitivity}
\vspace{6pt}
\setlength{\tabcolsep}{16.5pt}
\begin{tabular}{lcccccc}
\toprule
Method & 0.07 & 0.08 & 0.09 & 0.10 & 0.11 & 0.12 \\
\midrule
PBE        & 84.81  & 85.39  & 85.93  & 86.39  & 86.83  & 87.28  \\
HSE        & -64.00 & -64.18 & -64.35 & -64.47 & -64.55 & -64.62 \\
QP$GW$       & -94.60 & -94.61 & -94.62 & -94.65 & -94.67 & -94.70 \\
PBE+$U$(BO)  & -94.52 & -94.41 & -94.30 & -94.21 & -94.11 & -94.03 \\
\bottomrule
\end{tabular}
\end{table*}

\bibliography{reference}